\documentclass[prx,aps,twocolumn,reprint,longbibliography,floatfix]{revtex4-1}

\usepackage[utf8]{inputenc} 
\usepackage{graphicx}
\usepackage[colorinlistoftodos]{todonotes}
\usepackage{amsmath}
\usepackage{amssymb}
\usepackage{amsthm}
\usepackage{color}
\usepackage{color,soul}
\setstcolor{yellow}

\usepackage[modulo]{lineno}

\newcommand{\be}{\begin{equation}}
\newcommand{\ee}{\end{equation}}
\newcommand{\bea}{\begin{eqnarray}}
\newcommand{\eea}{\end{eqnarray}}

\begin{document}
\title{A Scanning Quantum Cryogenic Atom Microscope}
\author{Fan Yang}
\author{Alicia J. Koll\'{a}r}
\author{Stephen F. Taylor}
\author{Richard W. Turner}
\author{Benjamin L.~Lev}
\affiliation{Departments of Physics and Applied Physics and Ginzton Laboratory, Stanford University, Stanford, CA 94305}

\date{\today}

\begin{abstract}
Microscopic imaging of local magnetic fields provides a window into the organizing principles of complex and technologically relevant condensed matter materials.   However, a wide variety of intriguing strongly correlated and topologically nontrivial materials exhibit poorly understood phenomena outside the detection capability of  state-of-the-art high-sensitivity, high-resolution scanning probe magnetometers.    We introduce a quantum-noise-limited scanning probe magnetometer that can operate from room--to--cryogenic temperatures with unprecedented DC-field sensitivity and micron-scale resolution.  The Scanning Quantum Cryogenic Atom Microscope (SQCRAMscope) employs a magnetically levitated atomic Bose-Einstein condensate (BEC), thereby providing immunity to conductive and blackbody radiative heating. The SQCRAMscope has a field sensitivity of {1.4~nT per resolution-limited point ($\sim$2~$\mu$m), or 6~nT/$\sqrt{\text{Hz}}$ per point at its duty cycle.  Compared to point-by-point sensors,  the long length of the BEC provides a naturally parallel measurement, allowing one to measure nearly one-hundred points with an effective field sensitivity of 600~pT$/\sqrt{\text{Hz}}$ for each point during the same time as a point-by-point scanner would measure these points sequentially.}  Moreover, it has a noise floor of 300~pT and provides nearly two orders of magnitude improvement in magnetic flux sensitivity (down to $10^{-6}$~$\Phi_0/\sqrt{\text{Hz}}$) over previous atomic  probe  magnetometers capable of scanning near samples. These capabilities are, for the first time, carefully benchmarked by imaging  magnetic fields arising from microfabricated wire patterns, in a system where samples may be scanned, cryogenically cooled, and easily exchanged.   We anticipate the SQCRAMscope will provide charge transport images at temperatures from room--to--4~K in  unconventional superconductors and topologically nontrivial materials. 
\end{abstract}
\maketitle

\vspace{-3mm}
\section{Introduction}
\begin{figure}[ht!]\vspace{-0mm}
  \includegraphics[width=0.5\textwidth]{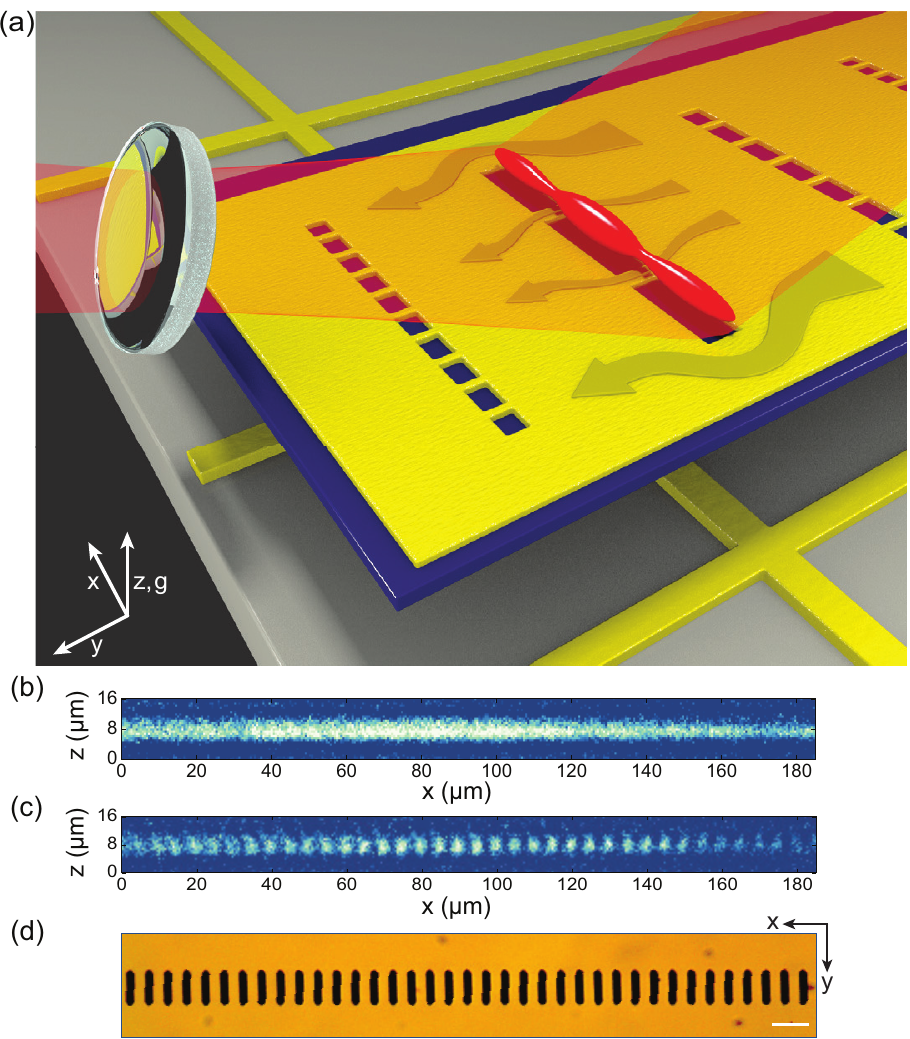}\vspace{-2mm}
\caption{SQCRAMscope operation:  (a), A quasi-1D BEC (red) is magnetically confined using an atom chip trap (light grey with gold wires). Suspended between the atom chip and the BEC is the silicon sample substrate (blue) onto which the gold  calibration pattern is fabricated. {The gaps in this gold define the microwires, and the magnetic field from current flowing through these wires (yellow arrows) fragments the BEC.}   Contact leads not shown. Atomic density is imaged with a high-NA lens by reflecting a resonant laser (transparent red) off the gold. BEC position is fixed while the sample substrate, not connected to the atom chip, may be scanned and cryogenically cooled.    For clarity, sketch is upside down with respect to gravity.  (b), Absorption image of an unfragmented quasi-1D BEC at a position $d=0.8(1)$~$\mu$m from the sample.  (No current flows.)  (c), Absorption image of fragmented BEC.  Current flows through periodic array of 2.5-$\mu$m wide wires spaced 2.5-$\mu$m apart.  (d), Image of the microwire array used in panels (b) and (c).   White scale bar is 10~$\mu$m.}
\label{fig1cartoon}
\end{figure}

\vspace{-3mm}
Quantum sensors comprised of nitrogen-vacancy (NV) color centers in diamond have joined scanning Superconducting Quantum Interference Devices (SQUIDs) in advancing high-sensitivity magnetometry into the nanoscale regime~\cite{Grinolds:2013gi,Vasyukov:2013ed,Kirtley:2016uj}.  {BECs have also been used for magnetometry}~\cite{Esteve:2004hj,Gunther:2005hm,Schmiedmayer05_Nature,Schmiedmayer06_APL,Aigner:2008}.
We add to this quantum metrology toolbox a carefully calibrated cryogenic scanning magnetometer that exploits the extreme sensitivity of {these} quantum gases to external fields.  The SQCRAMscope operating principle is sketched in Fig.~\ref{fig1cartoon}.  Inhomogeneous magnetic fields from a nearby source exert a Zeeman force on atoms Bose-condensed in a smoothly varying harmonic trap.  The atoms move in response, distorting the otherwise smooth wavefunction of the BEC.  The BEC density is then imaged by recording the absorption of resonant light using a CCD camera.  The local density may be related to field through the BEC equation of state.  A 2D-field map is created by raster-scanning the relative position of the BEC and the source with a duty cycle limited by the time needed to recreate the BEC after the destructive absorption imaging process.   Assuming no $z$-dependence of the source, application of the Biot-Savart law, conservation of current, and a  measurement of the distance $d$ between BEC and  sample allows one to convert an $\hat{x}$-$\hat{y}$ map of the $\hat{x}$ field component into a 2D map of the current flow (or magnetic domains) in the source~\cite{Esteve:2004hj,Gunther:2005hm,Schmiedmayer05_Nature,Schmiedmayer06_APL,Aigner:2008}.

\vspace{-0mm}

The BEC is confined in a high-aspect-ratio, cigar-shaped trap formed using an atom chip-based magnetic microtrap~\cite{fortaghrmp,naides2013trapping} (see Appendix~\ref{Exp}). This quasi-1D Bose gas lies within the quasicondensate regime because the transverse trap frequency $\omega_\perp$ is 157$\times$ larger than the longitudinal trap frequency $\omega_\parallel$, the chemical potential $\mu$ is similar in magnitude to $\omega_\perp$, and the gas temperature is 2.6$\times$ lower than $\omega_\perp$.  As such,  density fluctuations are suppressed below the quantum shot-noise limit~\cite{gerbier,Bloch2008,jacqmin}, enhancing field sensitivity (see Appendix~\ref{FieldCalib}).  The equation of state in the quasicondensate regime we employ is given by $\mu_\text{1D}(x)  =\mu - V(x) = \hbar \omega_\perp \sqrt{1 + 4 \ a\  n_\text{1D}(x) }$, where $a$ is the 3D $s$-wave scattering length,  $n_\text{1D}(x) $ is the line density, and $V(x)$ is the total potential~\footnote{We actually employ the full expression given in Ref.~\cite{gerbier}; see also Ref.~\cite{Kruger:2010eg}.}. This potential is the sum of the well-characterized trapping potential  and the magnetic potential $V_m(x)$ to be measured. For the employed atomic state and at the small fields that are measured, the magnetic potential is linearly proportional to field, i.e., $V_m(x)= \mu_B B_x(x)$.  {Small inhomogeneous fields primarily perturb the trap potential only along the weakly trapped axis $\hat{x}$ because $\omega_\parallel\ll\omega_\perp$. Thus, the atomic position can only move a minuscule amount in response to weak transverse fields. Imaging perturbations to the BEC density along $\hat{x}$ therefore provides a measurement of a vector component of the magnetic potential along $\hat{x}$.} {The spatial density modulation is measured by absorption imaging with a resonant laser at an intensity well above saturation and reflected at shallow angle from the reflective  sample surface. Much care is taken to account for the presence of the standing wave intensity pattern of the reflecting beam.  See Appendix}~\ref{ImagingCalib}.

The responsivity of the magnetometer is given by $\partial n_\text{1D}/ \partial B =  \mu_B^2 B/2 a \hbar^2\omega_\perp^2$. In the limit of low atom number, the equation of state can be approximated by $\mu - V(x) = 2 \hbar\omega_\perp a  n_\text{1D}(x) $, and the responsivity becomes $R=\mu_B/U_\text{1D}$, where $U_\text{1D}$ is the effective 1D-interaction strength $2\hbar\omega_\perp a$. We employ two different traps in this work, one whose trap frequencies are optimized for high sensitivity, and the other for extended dynamic range.  

{In the following, we benchmark a number of the SQCRAMscope's attributes.  These include: 1) The responsivity $R$ [nT/(atom/$\mu$m)] of the atomic density to magnetic field variations. 2) The field sensitivity (minimum detectable field). This is quoted in several forms depending on operation modality:  field sensitivity [nT] in a single-shot per resolution-limited point size (RLP), field sensitivity [nT/Hz$^{1/2}$] per RLP, and field sensitivity [nT/Hz$^{1/2}$] over a finite scan area (to be explained below).  3) Magnetic flux sensitivity [$\Phi_0$/Hz$^{1/2}$] both per RLP  and over a finite scan area, where $\Phi_0$ is the magnetic flux quantum.  4) The magnetic field noise floor [pT] via Allan deviation measurements, which determines the minimum detectable field provided by averaging. 5) Spatial resolutions [$\mu$m] of both the field at the BEC and the current in the sample a distance $d$ away. These resolutions are given according to the Rayleigh criterion, which determines the RLP. 6) Current density resolution [nA/$\mu$m], which is the current density resolvable by measuring the field a distance $d$ from the surface. 7) The accuracy [\%], which measures the ability to determine a known field within a certain percentage error. 8) The repeatability [nT] quantifies the ability to measure the same field in successive runs and is defined as the average standard deviation in the data. 9) The dynamic range  [$\pm$nT] is the range in field in which the sensor has a linear response $R$.} 

{To determine these specifications, several other quantities required careful calibrations, including the absorption imaging magnification, the BEC--to--surface distance $d$, the per-pixel photon shot noise at our CCD camera, the atomic density noise per pixel, and contributions from patch fields.   Information regarding these are reported in the appendices, along with details of the BEC production, the calibration sample, imaging theory, and cryogenic scan results. }

{The SQCRAMscope will enable the high-sensitivity study of strongly correlated and topologically nontrivial materials in unexplored regimes of high temperature and low frequency.   For example, domain structure and transport near twin boundary interfaces in underdoped Fe-arsinide superconductors can be explored as the  $T_{N}\simeq50$--150 K nematic transition is crossed}~\cite{Chu:2010cr}.  {Other unconventional superconductors and complex oxide interfaces (e.g., LAO/STO)}~\cite{Bert:2011cg} {may be explored at high temperatures, as well as the metal-to-insulator transition in VO$_2$}~\cite{Qazilbash:2007cm}.  {Both transport and static magnetization at the $\agt$100-K ferromagnetic-metallic and antiferromagnetic-insulating transition in colossal magnetoresistive systems may be imaged.   Investigations of topologically protected transport should be possible}~\cite{Dellabetta:2012fc,Nowack:2013it}, {as should investigation of the electronic hydrodynamic flow in graphene above LN$_2$ temperatures}~\cite{Zaanen:2016hn}. { Lastly, the SQCRAMscope will also find use in engineering hybrid quantum systems, in coupling BECs to photonic topological metamaterials, and in studying the Casimir-Polder force}~\cite{Lin:2004eya,Obrecht:2007ep}.

\begin{figure}[t!]
\includegraphics[width=0.52\textwidth]{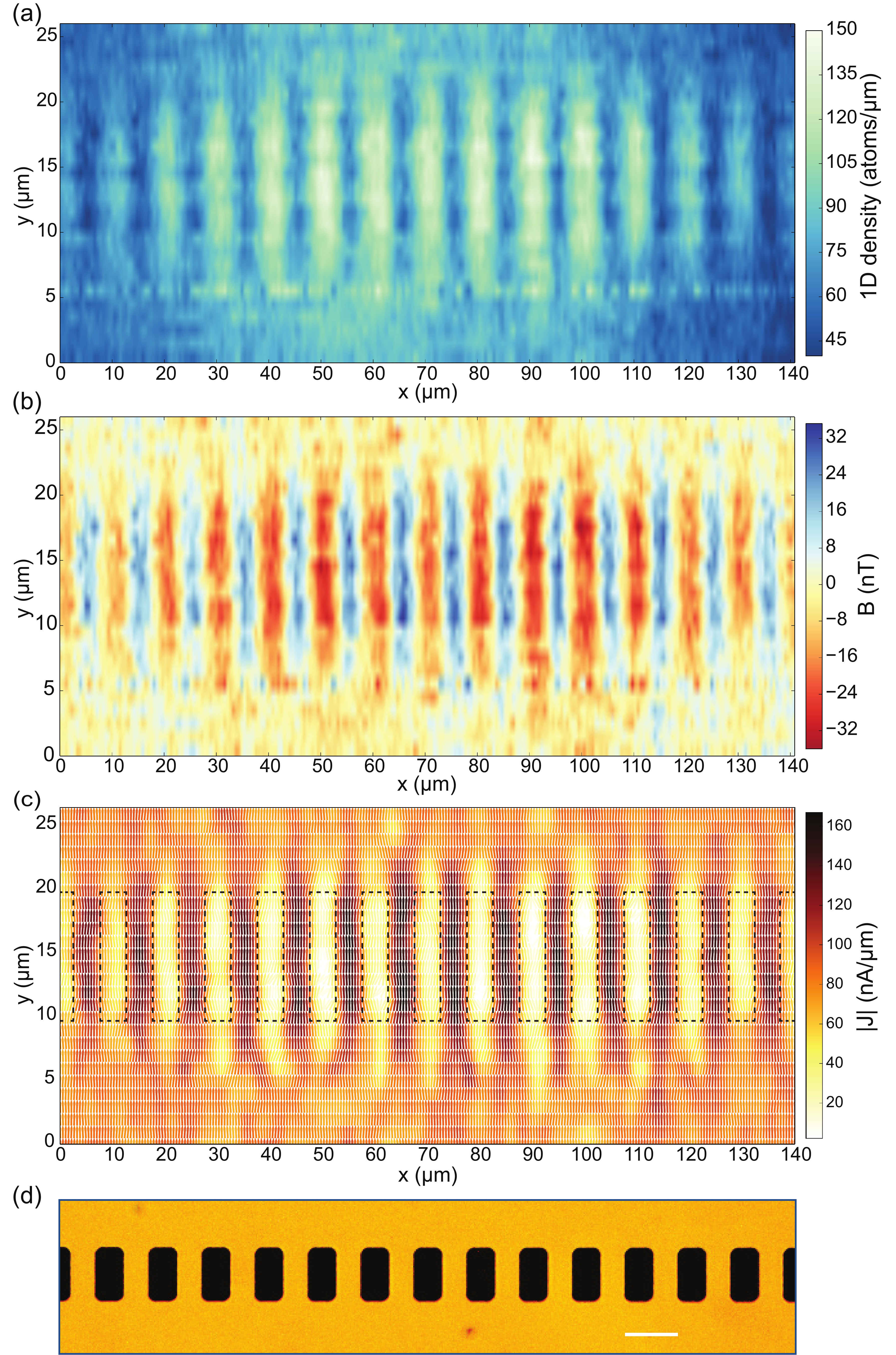}
\caption{Wide-area image of current density in microwire array. Current flows through room-temperature array of  5-$\mu$m-wide gold wires spaced 5-$\mu$m apart.  Sample is scanned in 1-$\mu$m steps along $\hat{y}$.  The BEC is confined 1.4(1)-$\mu$m below using the extended-dynamic-range trap. (a) Line density of BEC.  (b) Magnetic field along $\hat{x}$ derived from density data using equation of state. (c) Current density obtained from magnetic field and measurement of $d$ through use of Biot-Savart equation.  Arrows indicate current direction; black dashed rectangles demarcate the gaps between  microwires. {Current modulation away from wires is likely due to inhomogeneously resistive grain boundaries,} rather than noise, as explored in Ref.~\cite{Aigner:2008}. The spacing of arrows is given by our pixel size, $\sim$0.54~$\mu$m, while our spatial resolution for current flow is 2.6(1)~$\mu$m. (d) Image of the microwire array.  White scale bar is 10~$\mu$m.}
\label{fig2BECscan}
\end{figure}

\begin{figure}[h!]
\includegraphics[width=0.5\textwidth]{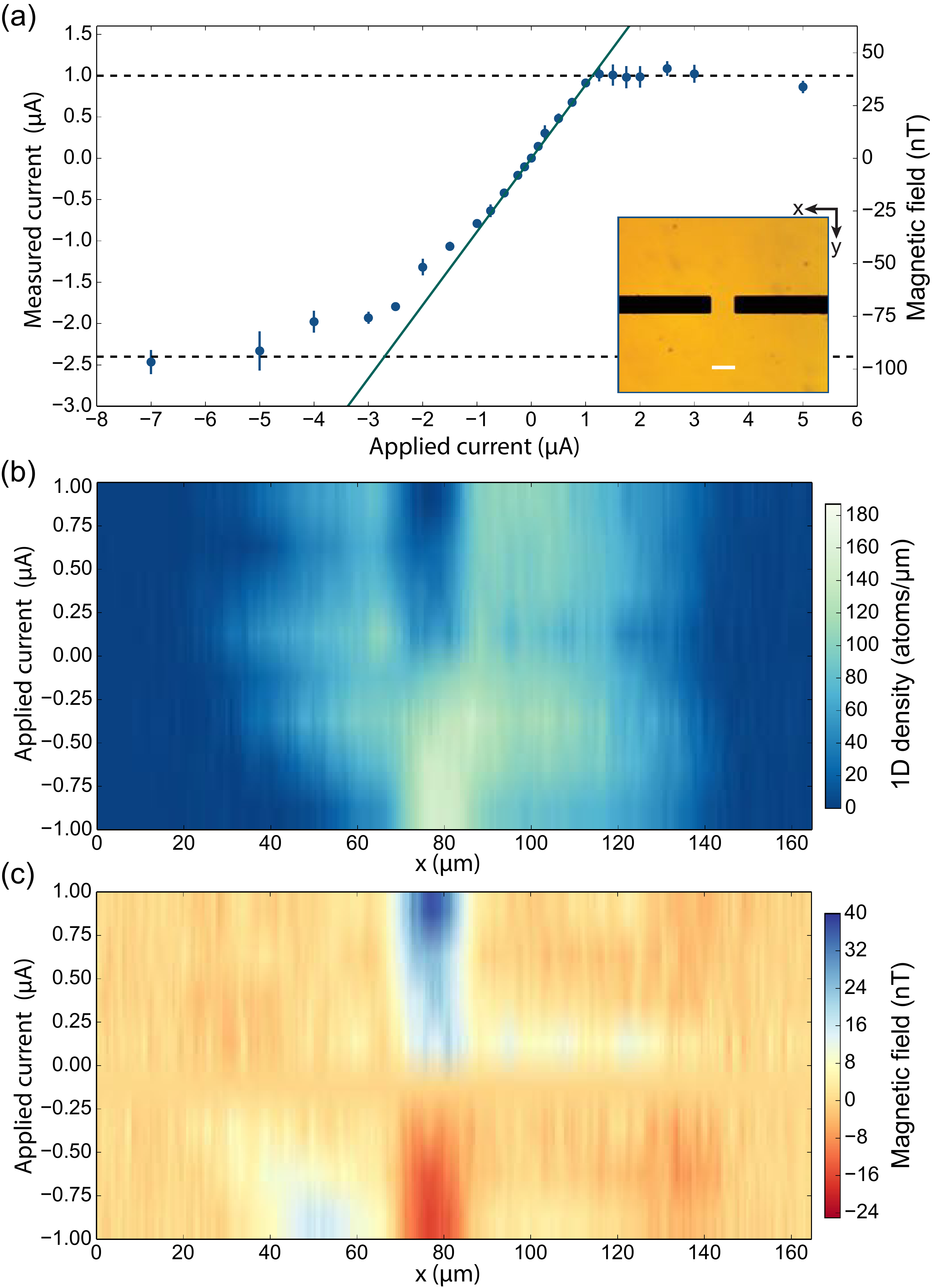}
\caption{ SQCRAMscope accuracy, repeatability, linearity, and dynamic range.   (a) Measurement of current in the 12-$\mu$m wide  calibration  wire shown in inset.  White scale bar is 12~$\mu$m.   BEC is positioned 1.7(1)-$\mu$m below  wire center using the high-sensitivity trap, and the current is varied to create either a density dimple or peak in the atomic density. The green line is a fit to the linear region.  (b) Atomic line density versus applied current used for the data in panel (a).  (c) Magnetic field in $\hat{x}$ calculated from data in panel (b).}
\label{fig3accuracy}
\end{figure}

\section{Experimental Results}

Figure~\ref{fig2BECscan} shows a typical magnetic field scan above a room-temperature microwire array, and the resulting image of the current density flowing through the wires. The FWHM point-spread resolution of the atomic density and magnetic field is 2.2(1)~$\mu$m, and can be improved in the future by more than a factor of two using a custom-tailored lens system.  {The current density map has a lower spatial resolution than the field maps, stemming from the convolution of the field resolution with the typical distance $d=1.4(1)$~$\mu$m between the atoms and sample.}  Distances as short as 0.8(1)~$\mu$m, shown in Fig.~\ref{fig1cartoon}(c), may be used, providing a source resolution of 2.3(1)~$\mu$m.   Current flow through the microwires is clearly visible in Fig.~\ref{fig2BECscan}(c), as is the fanning out of current into the bulk. Patches of lower current density may result from  higher-resistance grains in the polycrystalline film, as measured in Ref.~\cite{Aigner:2008}.    (See Appendices~\ref{Exp} and~\ref{ImagingCalib} for resolution measurements, distance calibration, and scan done with sample at cryogenic temperature.)

The accuracy, repeatability, linearity, and dynamic range of the magnetometer are shown in Fig.~\ref{fig3accuracy}.  The green line is a  fit to the linear region of the measured versus applied current data and has a slope of 0.88(2).  This implies an accuracy of 11(2)\% and a responsivity only 11(2)\% lower than that predicted   for the high-sensitivity trap, $R=1.97(3)$~nT/(atom/$\mu$m).  The fields and current densities reported in  Fig.~\ref{fig3accuracy} correspond to those calculated with a finite-element solver for the employed gold microwire dimensions to within $\sim$10\%.

The linear part of the dynamic range is between approximately $\pm1.0$~$\mu$A ($\pm40$~nT). The upper limit arises due to lack of atoms in the high-field regions, while the lower limit arises due to all the atoms pooling into the low-field regions, as may be seen in panel \textbf{b}.  The extended-dynamic-range trap provided a 3$\times$-larger linear dynamic range, though with a responsivity 3$\times$ worse (see Appendix~\ref{FieldCalib}).  We have employed a thermal gas to increase the dynamic range by $\sim$50$\times$, though at worse resolution and with $\sim$100-fold-worse responsivity, as also noted in Refs.~\cite{Schmiedmayer06_APL,Aigner:2008}. 

The repeatability, or average standard deviation in the data about the linear fit, is 2.3(3)~nT per pixel, and the stability of the field measurement, measured as Allan deviation, is 1.1(1)~nT after 30 experimental runs.  To calculate  repeatability and Allan deviation, we extract the (shot-to-shot) deviations of the measured total current in the 12~$\mu$m wire by averaging the difference between the measured field profile and the predicted field profile (of the 12~$\mu$m wire);  repeatability and Allan deviation are spatial averages over the width of the calibration wire.  

With no current applied to sample, we measure a single-shot minimum sensitivity of 2.8(5)~nT per pixel. (No spatial averaging is performed for this measurement.)  This value is  consistent with two independently measured quantum-noise-limited sources:  photon shot noise accounts for 2.5(4)~nT  per pixel, while atom density noise is 1.7(7)~nT  per pixel.  Due to the quasicondensate nature of the gas, this noise is a factor of $\sim$2 below that expected for shot noise  in the higher density regions of the BEC; hence the use of ``Quantum" in the name of the scanning microscope. (See Appendix~\ref{FieldCalib} for   Allan deviation and noise measurements.)  

The noise floor  is  measured also by Allan deviation,  and the result is $\sim$300 pT per pixel after 100 runs. (No spatial averaging is performed for this measurement.)   This noise floor  translates into a 2-nA minimum detectable current from an infinitely thin wire or 80 horizontally oriented 1-$\mu_B$ dipoles, when the BEC is located $d=0.8$~$\mu$m from these sources. 

 {Since the particular pixel sizes are not intrinsic to the sensor itself, we now convert these `per pixel' values to field sensitivities per our resolution-limited point (RLP) size of 2.2~$\mu$m. The RLP is set by our spatial resolution for field measurements.  There are 4.1(2) pixels per RLP for our lens system, and so the single-shot field sensitivity per RLP is a factor of $\sqrt{4.1}$ lower, or 1.4(1)~nT.  This is equivalent to 6.1(1)~nT/Hz$^{1/2}$ per RLP when considering the $\tau=16$~s duty cycle.}

\begin{figure}[t!]
 \includegraphics[width=0.48\textwidth]{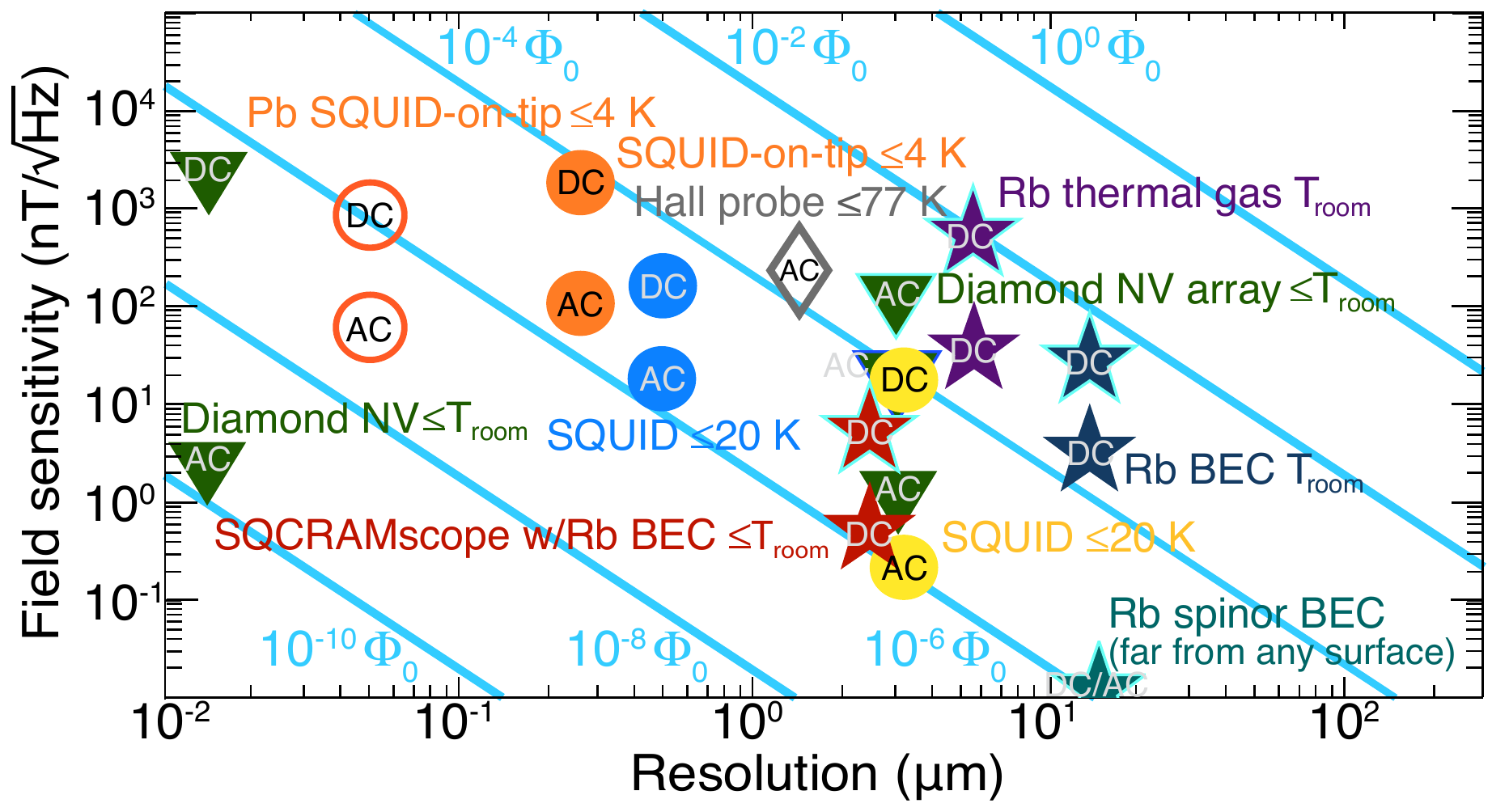}
\caption{Comparison to other scanning probes capable of imaging arbitrary field sources~\cite{Schmiedmayer05_Nature,Schmiedmayer06_APL,Aigner:2008,Kirtley:2012dx,Vasyukov:2013ed,Kirtley:2016uj,Iranmanesh:2014iu,Pham:2011dc,Grinolds:2013gi,Shields:2015ik,Lovchinsky:2016dqa,vengalattore,Brook2003}.  The diagonal cyan lines are contours of constant magnetic flux. Measurement bandwidth restrictions are labeled with ``AC" or ``DC".  Temperature range capabilities are indicated. {The star-shaped data markers are for BEC techniques}. {Points  with a light-blue outline denote single-point measurement sensitivity, while those without outlines (as well as the green-triangle data marker with the blue outline) represent the sensitivity for parallel, multi-point measurements; see text.}   Addition of a heat shield to the SQCRAMscope---red stars---will extend the temperature range from 35 K down to $\sim$4 K.  }
\label{compare}
\end{figure}

The magnetometer simultaneously provides $M$ independent RLPs of  information because the quasi-1D BEC may be several hundred microns in length while the imaging resolution is on the micron scale. {This provides an advantage over point-by-point scanning magnetometers such as SQUIDs because the SQCRAMscope can repeatedly measure these $M$ points during the time $MT$ it takes  the scanning SQUID to sequentially measure the $M$ points once. $T$ is the integration time per point.  That is, if $MT\gg \tau$, then  the SQCRAMscope has enough time to  average the signal of each RLP $MT/\tau$ times during the same time it would take the SQUID to record a single scan of the $M$ points.  This lowers the field sensitivity by a factor of $\sqrt{MT/\tau}$.  In terms of sensitivity per root Hz, the enhancement factor is $\sqrt{M}$.   This results in a field sensitivity of 590(80)~pT/$\sqrt{\text{Hz}}$ per RLP for a line scan, given our $M=200/2.2 = 91(5)$ and $\tau$.} Thus, high-sensitivity, high-resolution, and wide-area scans of condensed-matter samples can be made within a few hours.  Future improvements  can expedite this by reducing the duty cycle and elongating the usable BEC length.  {In} Fig.~\ref{compare}, {we include  this parallel measurement-based sensitivity in addition to the single-shot sensitivity to properly account for the  parallel-scanning advantage of our sensor:  that is, what one cares about for comparing a parallel scanning probe microscope is the sensitivity per point in the same time it would take a comparable point-by-point sensor to scan the same number of points.   }

\section{Comparison to other techniques}

We now highlight other demonstrated features of the SQCRAMscope.  Because the samples are physically detached from the chip,  we can rapidly replace the sample without disturbing the BEC production apparatus in {five days}.  This is much faster than the up to several months of current systems whose samples are attached directly to the atom chip.  
Secondly, the sample temperature may be independently controlled and stabilized, in contrast to samples attached to an atom chip, where trapping wire currents can uncontrollably heat the sample. In the case of superconducting atom chip experiments such as in Refs.~\cite{Nirrengarten:2006eh,Mukai:2007hh,Roux:2008hj,Muller:2010gw,reichel2010atom,Cano:2008df,Jessen:2013hw,Minniberger:2014ht,Chan:2014kb}, the sample temperature would be fixed to that of the superconducting chip and not tunable.
The SQCRAMscope allows any $\sim$1-cm$^2$ area, $\leq$150-$\mu$m-thin sample made of UHV-compatible material to be imaged from room--to--cryogenic temperatures. We have demonstrated here the functionality of the SQCRAMscope at room temperature and 35 K (see Appendix~\ref{Exp}) and believe operation down to $\sim$4~K will soon be possible with the addition of a heat shield~\cite{naides2013trapping}. Lastly, we mention that because what the SQCRAMscope measures are nT-strength, short-wavelength \textit{deviations} in the mean field along the trap axis, the microscope is insensitive to much  larger---up to hundreds of Gauss---background or long-wavelength fluctuating fields along this axis.  This feature obviates the need for careful magnetic field shielding of the apparatus.

In the following, we compare the scanning magnetometry capability of the SQCRAMscope to other  high-source-resolution, high-sensitivity scanning probe magnetometers relying on field sensing of arbitrary sources rather than magnetic resonance of spins. (For the latter, see Refs.~\cite{Rugar:2004bc,Lovchinsky:2016dqa} for nanoscale diamond NV and magnetic force microscope-based techniques).  See Fig.~\ref{compare}.   Imaging the Larmor precession  of spinor BECs has been used to measure  magnetic fields with high sensitivity~\cite{vengalattore}, though the technique has yet to be demonstrated near any surface and so the technique's sensitivity limit near a sample is unclear.  

By contrast, spin-polarized BECs have been used  to image the  field from and current flow through nearby room-temperature gold wires using the technique discussed here~\cite{Gunther:2005hm,Schmiedmayer05_Nature,Schmiedmayer06_APL,Aigner:2008,Ockeloen:2013je}. These works employed a BEC (thermal gas) $d=10$~(3.5)~$\mu$m from the room-temperature wires, resulting in a {field sensitivity  of  $\sim$4 (42) nT/$\sqrt{\text{Hz}}$ per RLP} at a current density resolution of 11~(5)~$\mu$m to simultaneously measure 67 (200) RLPs. We note that Refs.~\cite{Schmiedmayer05_Nature,Schmiedmayer06_APL,Aigner:2008} do not include extensive calibration data and so best-case estimates are used for comparison; i.e., we assume the reported resolution is FWHM rather than, e.g., 1-$\sigma$, and that the reported \textit{calculated} sensitivities are realized in their experiments (our calculated values are better than what we measure).  
As for our parallel sensor, the field sensitivity is obtained by multiplying their calculated minimum field by the square root of the ratio of duty cycle to number of simultaneously measured RLPs (obtained from Refs.~\cite{Wildermuth:2005vt,Aigner:2007uq}).  Again, the latter are defined as the usable length of the cloud  divided by the spatial resolution, and is included because this instrument is intended as a scanning probe and what matters is the entire time it takes to scan an area, not a point.   To facilitate comparisons to SQUIDs below, we estimate their  magnetic flux sensitivity to be $10^{-4}$~$\Phi_0/\sqrt{\text{Hz}}$ by multiplying the minimum field sensitivity by the square of the current density resolution.  

By comparison, our present work has lowered the minimum achieved field sensitivity to 0.6(1)~nT$/\sqrt{\text{Hz}}$ per RLP for simultaneously measuring 91 points at a superior resolution of 2.3(1)~$\mu$m (Rayleigh criterion).  Moreover, we rigorously calibrated this from the use of a known test source in the form of gold microwires at both room and cryogenic temperatures.  This  results in nearly a 100-fold magnetic flux sensitivity improvement to $10^{-6}$~$\Phi_0/\sqrt{\text{Hz}}$. Lastly, unlike these previous experiments, our result derives from a true sample, one not part of the atom chip itself and so may be scanned for wide-area imaging, cryogenically cooled (to a temperature different than the chip temperature), and easily exchanged.   All these features make our SQCRAMscope a uniquely versatile  scanning probe atom microscope for condensed matter materials.  

The point-by-point scanning technique of SQUID magnetometers~\cite{Kirtley:2012dx,Vasyukov:2013ed,Kirtley:2016uj}  provides high-AC-sensitivity $10^{-6}$~$\Phi_0/\sqrt{\text{Hz}}$ (10--100$\times$ worse at DC) imaging at length scales from a few-microns~\cite{Kirtley:2012dx} to $\sim$100-nm~\cite{Vasyukov:2013ed,Kirtley:2016uj} and down to dilution refrigerator temperatures.  (New SQUIDs-on-a-tip have achieved AC flux sensitivities of below $10^{-7}$ ~$\Phi_0/\sqrt{\text{Hz}}$ (DC, $10^{-6}$ ~$\Phi_0/\sqrt{\text{Hz}}$)~\cite{Vasyukov:2013ed}, though with the use of fragile Pb-based superconductors.)
The superior low-temperature and high-frequency imaging abilities of these scanning SQUIDs is complemented by the higher DC sensitivity and high-temperature compatibility of the SQCRAMscope. Specifically, these high-resolution SQUIDs lose sensitivity above sample temperatures of $\sim$20~K due to weak thermal links to the sample,  bringing the SQUID close to its superconducting critical temperature~\cite{Iranmanesh:2014iu}. (High-T$_c$ SQUIDs provide less sensitivity.)  Moreover, SQUID sensitivity decreases by roughly two orders of magnitude at DC due to 1/$f$ noise below $\sim$100~Hz.  We also note that while SQUIDs must often employ large fields to function, the SQCRAMscope is minimally invasive, because the field at the trap bottom---and at the nearby sample---can be as small as $\sim$1~G. Conversely, however, our current SQCRAMscope is incompatible with the application of large fields perpendicular to the sample, though fields up to nearly a kG can be applied along the trap axis, in the plane of the  sample. 
 
Nanoscale scanning NV diamond magnetometers are also compatible with samples from room--to--cryogenic temperatures~\cite{Pelliccione:2016de}, but are additionally operable outside ultrahigh vacuum.   They provide a  field sensitivity of a few nT/$\sqrt{\text{Hz}}$ with resolution down to a few nm near surfaces at frequencies above 10~kHz~\cite{Grinolds:2013gi,Shields:2015ik,Lovchinsky:2016dqa} and $\sim$2-$\mu$T$/\sqrt{\text{Hz}}$ at DC, with more than ten-fold improvements predicted~\cite{Taylor:2008cp}.  {Arrays of NV sensors have been used for parallel, wide-area imaging of AC fields}~\cite{Pham:2011dc}. {In }Fig.~\ref{compare}, {the three green-triangle markers denote the sensitivity and resolution for these NV arrays for single-point measurements (light-blue outline), at higher effective AC sensitivity ($\sim$15~nT/$\sqrt{\text{Hz}}$ per RLP) with respect to the time it would take to integrate the same area as the raster line-scanned SQCRAMcope (shown with blue outline), and at still higher effective AC sensitivity ($\sim$1~nT/$\sqrt{\text{Hz}}$ per RLP) with respect to point-by-point sensors (no outline).}  The SQCRAMscope complements this probe with its nearly 1000-fold higher DC-sensitivity to fields that cannot be rapidly modulated.  {While the present $\sim$2-micron-scale SQCRAMscope resolution is far worse than single-NV probes,}  a SQCRAMscope with sub-micron resolution is possible as well as with AC sensitivity at bandwidths up to $\sim$10~kHz via dispersive imaging~\cite{vengalattore} and up to $\sim$MHz using Rabi spectroscopy~\cite{Treutlein04}.

\section{Outlook}

 {The unprecedentedly low DC-field sensitivity and wide temperature compatibility of the SQCRAMscope will provide the ability to investigate phenomena in materials outside the reach of the  current capabilities of, e.g., scanning SQUIDs.  We highlight here one such  application for which the SQCRAMscope is uniquely suited.  The possible emergence of the unconventional pair-density-wave superfluid state in cuprate, high-T$_\text{c}$ superconductors and the theory of striped superconductors is the subject of much interest and speculation}~\cite{Berg:2007kn,Berg:2009cr}.  {A key signature of such a state would be the spontaneous generation of DC currents around domains of the emerging chiral superconductor as the temperature is tuned below $\sim$30 K.    Imaging  transport may reveal these spontaneously generated current loops.  However, while the length scale of current loops may be on the few micron scale and thus within our detection range, it is not theoretically known what might be the magnitude of the currents and detection has been elusive}~\cite{Kivelson16}.  {Thus, one requires a sensor of highest DC sensitivity to maximize one's chances of observation and to bound this magnitude if no signal is observed.  The SQCRAMscope is uniquely positioned to tackle this important measurement:  its DC sensitivity is two--to--three orders of magnitude greater than SQUIDs and NV diamonds, and scanning a wide field-of-view will be necessary to hunt for the chiral currents as they emerge, a task appropriate for the SQCRAMscope with its parallel sensing capability. Moreover, the high AC sensitivity of those sensors is not relevant since the spontaneously generated currents cannot be modulated, and SQUIDs become less sensitive at the elevated temperature at which the transition is predicted to occur. In summary, the SQCRAMscope will enable the study of strongly correlated and topologically nontrivial materials in regimes of parameter space unamenable to any other current technique.}

\begin{acknowledgements}
We thank M.~Naides, J. DiSciacca, S. Qiao, and J. Straquadine for experimental assistance, and P.~Abbamonte, J. Analytis, S.~L.~Cooper, I.~Fisher, J.~Kirtley, S.~Kivelson, M.~Lukin, K.~Moler, R.~Walsworth, and A.~Yacoby  for stimulating discussions.  This work was supported by the U.S.~Department of Energy (DOE), Office of Science, Basic Energy Sciences (BES), under Award \#DE-SC0012338 (equipment and materials, and support for B.L.L.).  R.W.T., A.J.K.~and B.L.L.~acknowledge support from the Gordon and Betty Moore Foundation through Grant GBMF3502.
\end{acknowledgements}

\appendix

\section{Experimental Methods}\label{Exp}

\subsection{Cryogenic scan}

\begin{figure}[t!]
\includegraphics[width=0.5\textwidth]{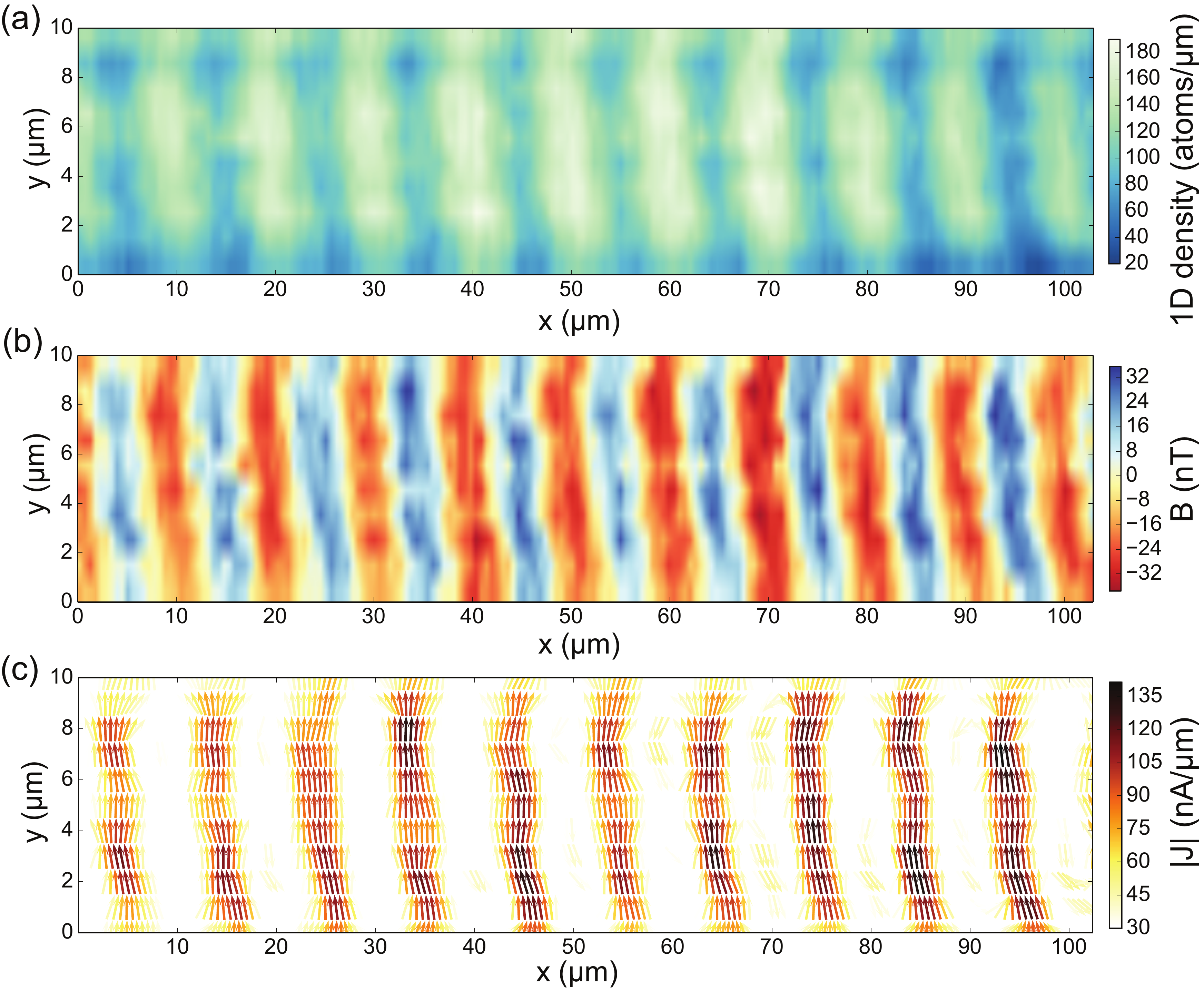}
\caption{Scan under cryogenic conditions. Sample at 35 K.  Scan over central region of the same microwire array as in Fig.~\ref{fig2BECscan}.  Wires are 5-$\mu$m wide and spaced 5-$\mu$m apart. Sample is moved in 1-$\mu$m steps along $\hat{y}$. The BEC is confined 1.2(1)-$\mu$m below it using the extended-dynamic-range trap. (a) Line density of atoms.  (b) Magnetic field along $\hat{x}$. (c) Current density. Arrows show direction of current flow.  The spacing of arrows is given by our pixel size, $\sim$0.54~$\mu$m, while our spatial resolution for current flow is 2.5(1)~$\mu$m.}
\label{Cryoscan}
\end{figure}

The cryogenic capability of the trapping apparatus was demonstrated in Ref.~\cite{naides2013trapping}.  While the majority of the data shown here were taken at room temperature, we performed a magnetometry scan, shown in Fig.~\ref{Cryoscan}, at 35~K with no discernible loss in capability, other than the observation of a horizontal drift in the relative BEC--sample position.  This is likely due to the slow settling of thermally contracting parts of the apparatus, and can be mitigated by waiting longer before imaging. Temperature is measured with in-vacuum thermometers contacted to the cold head and calibrated to the temperature at the sample tip~\cite{naides2013trapping}. 

\begin{figure*}
 \includegraphics[width=0.95\textwidth]{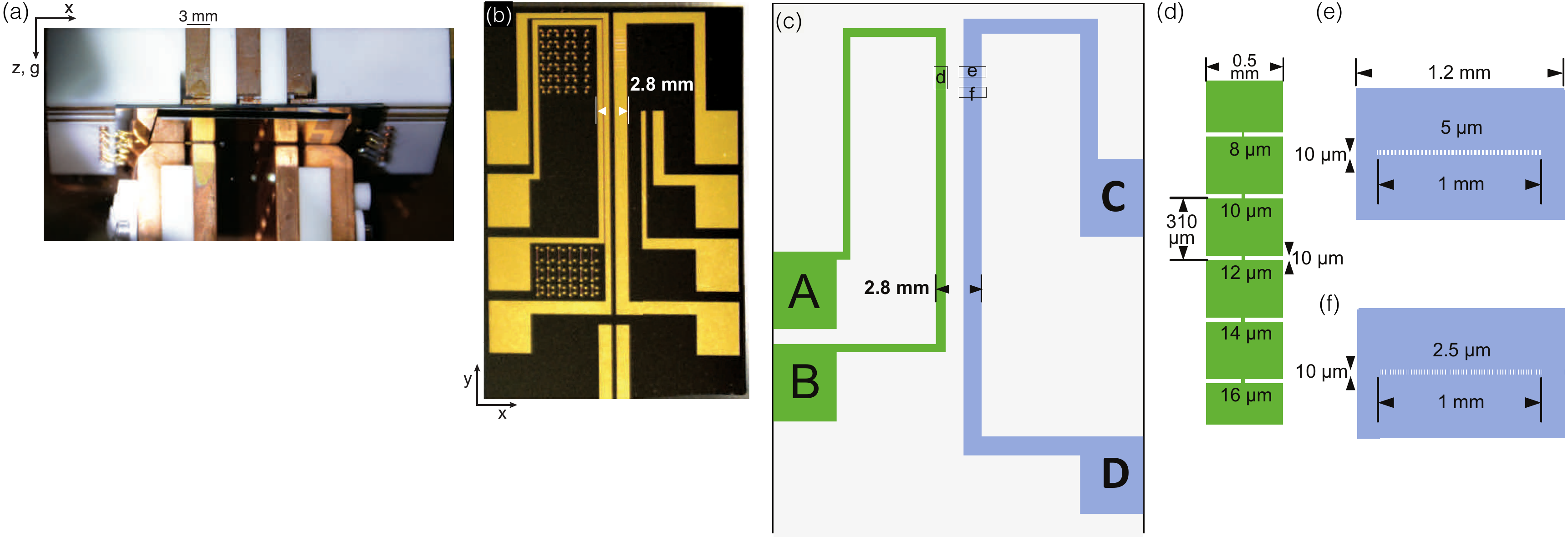}
\caption{Calibration sample schematic. (a) Image of gold-coated silicon sample substrate 200-$\mu$m below atom chip.  Copper leads for macrowire trap may be seen above atom chip, and trench through which the imaging beam passes is visible below the center of the sample substrate.  (b) Picture of  calibration sample.  (c) Schematic of the subset of wires in the calibration sample used for this work.  The green wire between contacts A and B supports the single microwire features shown in zoom in (d) while the blue wire between contacts C and D supports the arrays of microwires, two of which are shown in (e) and  (f).  Letters surrounded by black boxes denote the wire regions sketched in panels (d)--(f).  (d) Schematic of single microwire features, with the 12-$\mu$m-wide feature used for the data in Figs.~\ref{fig3accuracy},~\ref{EDRaccuracy}, and~\ref{AllanDev}.
 (e) Schematic of periodic array of 5-$\mu$m-wide microwires spaced  5 $\mu$m from one another.  This feature is used for the wide-area scans in Figs.~\ref{fig2BECscan} and~\ref{Cryoscan}.  (f) Schematic of periodic array of 2.5-$\mu$m-wide microwires spaced  2.5~$\mu$m from one another.  This feature is used for the image in Fig.~\ref{fig1cartoon}(c) and the resolution data in Fig.~\ref{Resolution_5um}.}
\label{Sample}
\end{figure*}

\subsection{Calibration sample}

The sample substrate is a 150-$\mu$m thick wafer of oxidized intrinsic $\left<100\right>$ silicon.
{The gold calibration wires are fabricated directly  onto the sample substrate using photolithography,} as shown in Fig.~\ref{Sample}(a),  and positioned, using a three-axis translation stage, such that the calibration wires are $\sim$300~$\mu$m below the atom chip.  See Ref.~\cite{naides2013trapping} for more details.  We have verified that the quasi-1D BEC, located below the calibration wires, is not fragmented by meandering currents in the atom chip's microwires  at this far distance.    The stage can move the sample substrate 3.8~mm in $\hat{x}$ and 5.4~mm in $\hat{y}$ allowing any sample feature within this area to be placed in proximity to the BEC. (Future improvements will increase this area by $\sim$5$\times$.) Coarse positioning of the BEC with respect to the microwire arrays on the calibration sample is made by fluorescence imaging from below the sample.  Fine positioning is performed by fitting magnetic field profiles obtained from magnetometry of current flowing through sample calibration features.   Microwire centers are found to within 1~$\mu$m.  

Vertical vibrations of the cantilevered calibration sample at frequencies larger than $\sim$1 Hz  are $\sim$150 nm RMS, much less than our imaging resolution, and may  be ignored~\cite{naides2013trapping}.  Thermal expansion of the atom chip and the sample mount assemblies cause small drifts of sample position relative to the BEC on a time scale much slower than the duty cycle.  However, these are easily measured via imaging and cancelled via feedforward to the bias field that controls the atom chip trap height between each run.   More difficult to cancel are vertical drifts at the time scale of the duty cycle.  These cause $\sim$0.5-$\mu$m shot-to-shot fluctuations of the BEC position $d$ relative to the sample.  Fortunately, these may be accounted for in the data analysis on a  shot-to-shot basis since  $d$ is imagined at the same time as the atomic density. The BEC scan data and accuracy calibrations reveal that the horizontal movements of sample stage with respect to the imaging system is less than 0.5~$\mu$m shot-to-shot at room temperature.  

The  data for Figs.~\ref{fig1cartoon}--\ref{fig3accuracy} were taken using the sample shown in Fig.~\ref{Sample}(b). We used e-beam evaporation, photolithography, and ion-milling to fabricate the 400-nm-thick gold calibration wire patterns.  The calibration sample contains two 1.2-mm-wide strips of gold, as sketched in Fig.~\ref{Sample}(c).  Gaps in the gold of these strips define microwires and  microwire arrays of different sizes and spacings, as shown in Fig.~\ref{Sample}(d)--(f). Current flowing through the constricted regions provides the signal for all measurements. The microwire arrays in panels \textbf{e} and \textbf{f} are periodic with equal wire width and spacing.  The magnification calibration in Fig.~\ref{Magnification} and resolution calibration in Fig.~\ref{Resolution_2um}(b)-(d) was taken with a different calibration sample (not shown), which contained an array of microwires of width 2.0(1)~$\mu$m and length 10~$\mu$m spaced by 20.0(2)~$\mu$m from each other.

\subsection{Quasi-1D BEC production} Atom chip trapping of ultracold gases of $^{87}$Rb atoms proceeds in a manner similar to that described in Ref.~\cite{naides2013trapping}.  Briefly, $4\times10^7$ atoms in the $|F,m_F\rangle = |2,2\rangle$ ground state at 30~$\mu$K are loaded into an optical dipole trap (ODT).  (The $g$-factor is $1/2$ in this weak-field seeking state.)  These atoms are then transported 33~cm by moving the lens focusing the ODT.   The atoms pass through a gate valve  into a science chamber.  The gate valve allows samples within the science chamber to be exchanged without breaking the ultrahigh vacuum of the production chamber.  The room-temperature vacuum of the science chamber is below $6.5\times10^{-11}$ Torr, sufficient for BEC production using the atom chip; see Refs.~\cite{fortaghrmp,reichel2010atom} for details on BEC production using atom chips.  This pressure decreases upon the cryogenic cooling of the sample~\cite{naides2013trapping}.   

The ODT is located 3.5-mm below the atom chip so as not to heat the sample and chip with scattered light.  To bring the atoms closer to the sample, the atoms are  transferred from the ODT to a magnetostatic harmonic trap formed by an H-shaped Cu-wire and an homogeneous field. The wire is mounted just above the atom chip, which faces downwards.
See Fig.~\ref{Sample}(a) and Ref.~\cite{naides2013trapping}.  This macrowire magnetic trap is moved upwards  to within a few hundred microns of the chip. Atoms are then transferred to an atom-chip-based trap.  The trap in the $\hat{y}$-$\hat{z}$ plane is formed with three wires:  a bias magnetic field along $\hat{y}$ is created by macrowires on either side of a 150-$\mu$m-wide wire whose current flows opposite to the macrowires; all current in these wires can be rapidly shut off to drop the atoms for short time-of-flight (TOF) imaging.   Weak confinement along the axis of the quasi-1D BEC is provided by  independently controlled orthogonal end-cap wires.  RF evaporation produces BECs with up to a few $10^4$ atoms.  The entire procedure is typically repeated every 20~s, though 16-s cycle times can be used.  

We create BECs in two different traps. The first, the high-sensitivity trap, is optimized to minimize single-shot repeatability and  detectable field.  (Lower transverse trap frequency leads to higher sensitivity through the minimization of mean-field energy, but also reduces the dynamic range due to lower density.) The trap frequencies are $\omega_\perp = 2\pi \times 710(10)$~Hz and $\omega_x = 2\pi \times 4.48(3)$~Hz, and we create quasi-1D BECs of $7.0(4)\times10^{3}$ atoms at temperature 12(1) nK.  (Larger BECs are possible, but the lower population ensures operation within the quasicondensate regime.)  This temperature is far below the quasidegeneracy temperature 1.5(1) $\mu$K---the critical temperature for phase fluctuations is 9.0(5)~nK---and 2.8$\times$-below the temperature associated with $\omega_\perp$, 33.7(6)~nK~\cite{gerbier,Bloch2008,jacqmin}.  (See Appendix~\ref{FieldCalib} for discussion of temperature measurement.) The chemical potential is  32(1)~nK, similar to $\omega_\perp$ and sufficiently low for the quasicondensate equation of state to hold to high accuracy~\cite{gerbier}. Shot-to-shot temperature and number fluctuations do not affect the magnetometry because they are not sufficiently large to change the applicable equation of state for the gas and are recorded and accounted for on a shot-by-shot basis. The second trap, with  frequencies $\omega_\perp = 2\pi \times 1810(30)$~Hz and $\omega_x = 2\pi \times 4.1(1)$~Hz, is optimized to extend the dynamic range while also providing 1D-Bose gases within the quasicondensate regime.

After  quasicondensate preparation, the current in the calibration wire is adiabatically ramped-up in 300~ms.  The BEC is then raised to a distance 1.7(2)-$\mu$m below the sample surface for the high-sensitivity trap and 1.3(4)-$\mu$m below the sample for the extended dynamic range trap.  Lifetime within the trap is greater than 700~ms at $d > 1$~$\mu$m; the attractive Casimir-Polder potential does not pose a severe lifetime limit at this distance~\cite{Lin:2004eya}.

\section{Imaging Calibration}\label{ImagingCalib}

 \begin{figure}[t!]
 \includegraphics[width=0.48\textwidth]{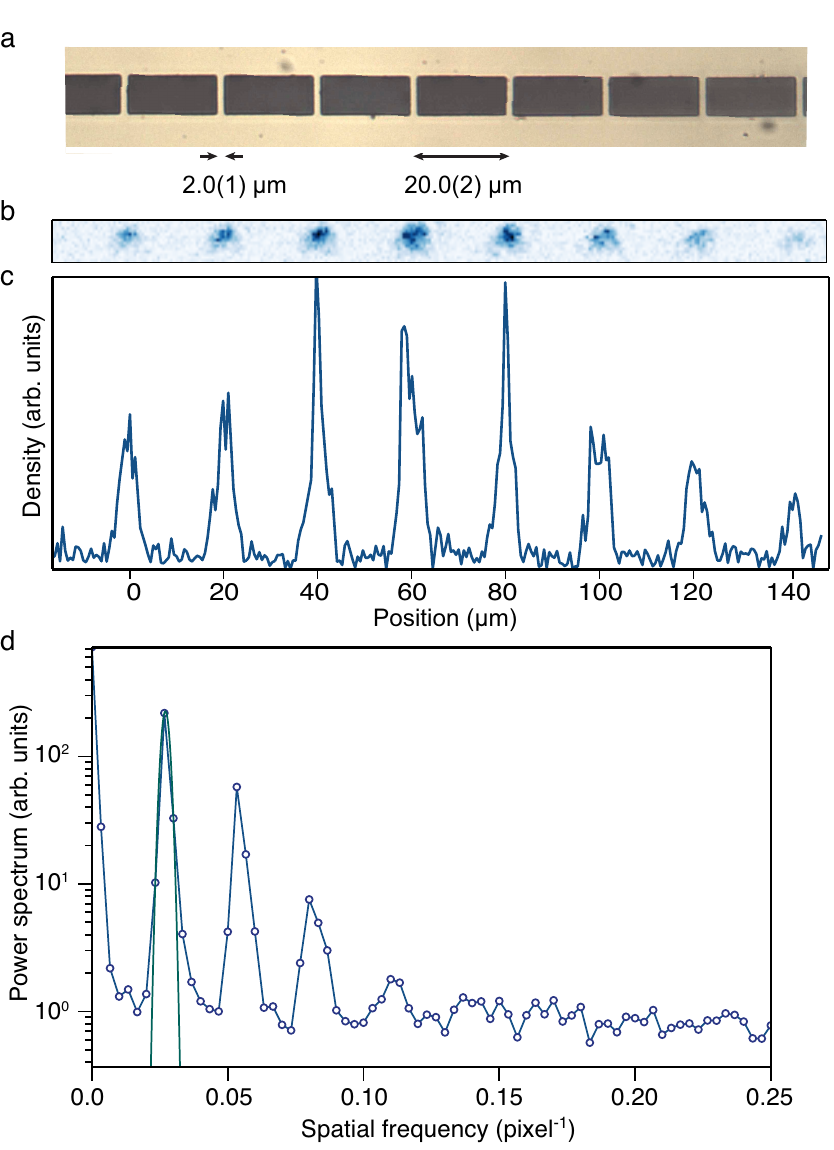}
\caption{Magnification calibration.  (a)  Image of calibration wire array with 2.0(1)-$\mu$m-wide wires spaced 20.0(2)~$\mu$m apart.  (b) Absorption image of BEC fragmentation, and (c)  density of atoms versus position.  (d) Power spectrum versus spatial frequency.  Green curve is fit to fundamental harmonic, from which we determine a magnification of 24.0(2).  This is consistent with  the  geometrically determined value of $\sim$24. Blue lines connecting data points are guides to the eye.}
\label{Magnification}
\end{figure}

\subsection{Imaging}
BECs are held in the magnetic field of the calibration sample for $\sim$250~ms before the trap and sample fields are rapidly removed for BEC imaging. 
This is much longer than the density response time of $\sim$0.5~ms, the time atoms take to move $\sim$1~$\mu$m at the speed of sound in the BEC~\cite{ketterle:1999de}.
We allow the atoms to fall for 150~$\mu$s.  The imaging beam is on for 20~$\mu$s, during which the atoms diffuse by $\sim$500-nm RMS, much less than our imaging resolution (see below).  During this brief TOF, the atoms fall $\sim$1 $\mu$m due to gravity and an initial velocity  imparted from magnetic gradients when the trapping fields are turned off.  The imaging takes place when the atoms are $\sim$2.5 $\mu$m from the sample (see below).  Ballistic expansion during this time is approximately 200~nm and also small compared to the imaging resolution.  

Accurate measurement of the imaging magnification via rate-of-fall-under-gravity is problematic due to the small field-of-view in the vertical direction.  This is the result of our high magnification and the use of only a small fraction of our CCD array so that fast kinetics mode can be used for fringe suppression.  We therefore measure the magnification  by imprinting a density modulation of known periodicity onto the gas and imaging its density in situ with no TOF. See Fig.~\ref{Magnification}.  The density  modulation is created by using the magnetic fields from a microfabricated microwire array on a calibration sample. Error in the spatial frequency measurement, combined with  uncertainty in the microwire dimensions, leads to a 0.8\% uncertainty in magnification.

\begin{figure}[t]
 \includegraphics[width=0.48\textwidth]{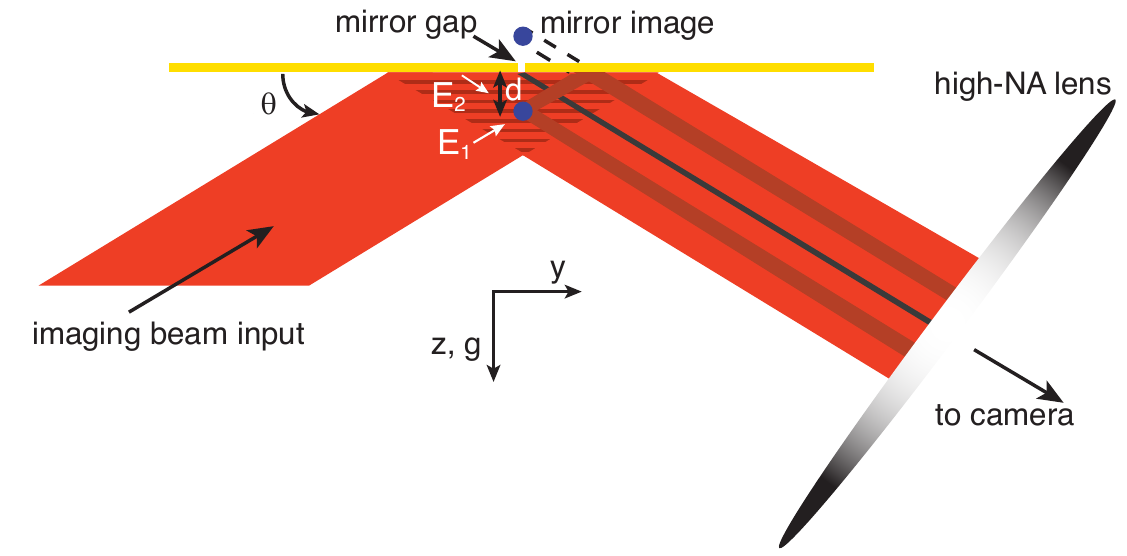}
\caption{ Geometry of absorption imaging beam reflecting off gold mirror of calibration sample.  (Not to scale.)  Gaps in the gold mirror, to define the microwires, cast shadows onto the reflected beam.   By design, however, the height of the shadow at the position of the BEC is much less than $d$. Non-reflective samples may be coated with a thin insulator before mirroring with gold; a gold mirror and insulating layer  need only be a few hundred nm, far thinner than our optical resolution.}
\label{Imaging}
\end{figure}

Imaging the BEC near a surface is complicated by the presence of diffraction fringes from the sample edge.  Reflecting the absorption imaging beam at a small angle $\theta$ removes the fringes from the field of interest \cite{schmiedmayer_imaging}, but it introduces a standing wave intensity pattern perpendicular to the chip.  (Non-reflective samples may be coated with a thin insulator before mirroring with gold; a gold mirror and insulating layer  need only be a few hundred nm, far thinner than our optical resolution.)    This  complicates the usually simple technique of absorption imaging commonly employed in free space, away from reflective surfaces. Moreover, due to the reflection, the polarization of the imaging beam must be linear and parallel to the sample surface to prevent polarization imperfections at the position of the atoms.  To define the quantization axis during imaging, we maintain a small B-field along this polarization direction.  The light drives $\pi$-transitions, resulting in non-closed-cycle transitions.  However, we have used numerical simulations of the optical Bloch equations to account for the optical pumping out of the trapped $m_F=2$ state during the imaging time; we use in our calculations the effective cross section $\sigma_0$ obtained from these simulations.  (This also accounts for the $\alpha^*$ term in Ref.~\cite{Reinaudi2007}.) We now describe how to perform absorption imaging in the presence of the standing wave pattern.  Fig.~\ref{Imaging} depicts the imaging geometry.  

The main points that must be considered are:  1) The atoms appear in two images on the CCD, a mirror image  from field $\vec{E}_1$, and a real image from the reflected field $\vec{E}_2$, as shown in  Fig.~\ref{Imaging}.   By mirror image we mean that the light scatters first from the atoms before  reflecting off the mirror into the imaging system. We use the average of the  mirror image and the real image for the data in this work.  2) The mirror is finite in extent and therefore acts as a Fourier filter for shallow-angle components that are not reflected into the imaging system~\cite{schmiedmayer_imaging}.  That is, the upward scattering fields from $E_1$ and $E_2$ are not imaged onto the CCD since their negative $k_z$  components do not get reflected from the mirrored sample.  While this can lead to image distortion, it is a negligible perturbation for mirrors as long and  $d$'s as small as ours.  3) Angular aliasing:  for narrow gases of atoms imaged at small $\theta$, light scattered away from the mirror at angles larger than $\theta$ will not be represented in the imaging system as having come from the actual gas location (see Ref.~\cite{ schmiedmayer_imaging}).  This effect blurs the lower half of the gas image in  $\hat{z}$ and the atom number can be miscounted for a gas much narrower than imaging resolution such as ours.  However,  this aliasing is not an issue for our system because for the high-saturation imaging we employ, summing over the apparent density in  $\hat{z}$ counts all the atoms due to the linearity of the solution to Beer's law in this limit (see below). 4) The standing wave intensity pattern changes the local rate of scattering.  For example, atoms located at the node of the standing wave do not register on the camera.  5) The interference pattern is not present at the CCD plane because $\vec{E}_1$ and $\vec{E}_2$ from the imaging beam are recorded on separate areas of the CCD.  As a consequence, there are field components recorded by the CCD camera that are not present at the atoms. 

The effects of points 4 and 5 are non-negligible and must be taken into account to obtain an accurate measurement of atom density.  We do so by modifying Beer's law for the case of absorption imaging an optically dense gas near a reflective surface with high optical resolution.  We present a thorough discussion of this derivation because it has not been presented in the literature, as far as we are aware, but is crucial to the high accuracy achieved by the SQCRAMscope.

In order to measure the atomic density, we take three images 1.5-ms apart.  In the first image, the atoms are illuminated by the probe beam  and this casts a shadow on the imaging system with intensity profile $P^F$.  The second image is taken without the atoms present, yielding $P^I$ at the imaging system.  Finally, a third background image is taken in the absence of both the probe beam and the atoms to measure any stray light incident on the camera. This background image is then subtracted from the first two images.  

Traditional absorption imaging is performed well below atomic saturation intensity, $I_{sat}$, so that power scattered by the atoms is proportional to the probe intensity. In this regime, the atomic scattering rate is given by $\rho \sigma_0 I$,  where $\sigma_0$ is the effective atomic cross section and $\rho(x,y,z)$ is the volume density of the cloud. This leads to Beer's law, a differential equation $dI /dy = - \rho \sigma_0 I$ that describes the evolution of the probe beam intensity.  It supports an exponentially decaying solution resulting in $n \sigma_0 =  -\ln (I^F/I^I) =  -\ln (P^F/P^I)$, where $n(x,z)$ is the atomic column density, $I^I(x,z)$ is the incoming intensity, and $I^F(x,z)$ is the intensity after passing through the gas.  For free-space imaging, the measured intensities are directly related to the intensities at the atomic position: $P^I = I^I$ and $P^F =I^F$. 

This simple, low-intensity imaging method, found sufficient for other atom chip systems~\cite{schmiedmayer_imaging}, cannot be used in our high-magnification, high-optical-depth system:  the large magnification of our high-resolution imaging system means that low probe intensities correspond to very low count levels on the camera, while the high optical density (OD) of our nearly in situ BEC means that almost all of these few photons are absorbed.  The atomic response is saturated, resulting in low dynamic range.  Detuning off resonance reduces OD, but at the expense of image distortions due to the  dispersive atomic medium.  
We therefore choose to operate in the opposite regime of large probe intensity to completely saturate the atomic response~\cite{Reinaudi2007}.
In this regime, a new term must be included in the equation for $n$. In the case of resonant imaging with a single traveling-wave beam in free space, the relation becomes 
\begin{equation}\label{BeersSolution}
n\sigma_0 = -\ln(I^F/I^I)+(I^I-I^F)/I_{sat}.
\end{equation}

For the probe intensities used here, the second, linear term is the dominant contribution, rather than the first, nonlinear term, as in low-intensity imaging.  This is  advantageous for systems, such as ours, in which large variations in the column density can occur on length scales smaller than the imaging resolution. Specifically, such effects cause  $P^F$ to differ from  $I^F$, with the result that Eq.~\ref{BeersSolution} does not yield the actual in situ density.  The discrepancy can be large in the $\hat{z}$ direction transverse to the BEC axis where the gas is smaller than the imaging resolution.  But since the measured magnetic field depends only on $n_{1D}(x)$, the integral along $z$ of the column density $n(x,z)$,  it is sufficient to determine the total column density integrated along the $\hat{z}$ direction.  That this integral accurately counts the atoms is due to the linearity of Eq.~\ref{BeersSolution} in the high-intensity imaging limit.

The intensity pattern at the atoms consists of two interfering traveling waves, $\vec{E}_1(x,y,z)$ due to the portion of the imaging beam that passes through the atoms before reflecting off the sample, and $\vec{E}_2(x,y,z)$ due to the portion that reflects before passing through the atoms (see Fig.~\ref{Imaging}).  Restricting our attention to the region of the interference pattern near the atoms and assuming equal intensities in both beams, the magnitudes are related by $|E_2(y,z)| = |E_1(y,z)| \equiv \mathcal{E}(y,z)$ in the limits that the angle of incidence to the sample is small and that the gas size is small compared to the wavelength of the standing wave.  These conditions are well satisfied in our system.  Specifically, the two fields take the form 
\bea
E_1(y,z) &=& \mathcal{E}(y,z) \exp{(i k_\parallel y + i k_\perp z)},  \\
E_2(y,z) &=& -\mathcal{E}(y,z) \exp{(i k_\parallel y - i k_\perp z)},
\eea where $k_\parallel$ is the projection of the wavevector onto the $y$ direction parallel to the sample,  $k_\perp$ is the projection of the wavevector onto the $z$ direction perpendicular to the sample, and the minus sign accounts for the phase shift imparted by reflection off the sample. The total electric field  takes the form 
\be
E (y,z) = 2i \mathcal{E}(y,z) \sin(k_\perp z) \exp({ik_\parallel y}),
\ee
which gives rise to the interference pattern in the intensity at the atoms of 
\be
I(y,z) = 4\mathcal{E}^2(y,z)\sin^2(k_\perp z).
\ee

To calculate the measured $P$ intensities from the $I$'s, we now must relate $|E_1|^2$,  which is the field that forms the upper, mirror   image on the camera, to $I$, which sets the atomic scattering rate. One must  contend with the fact that the two  fields $E_1$ and $E_2$  propagate in  different directions and so are imaged without interference at separate locations on the CCD.  The analog of Beer's law is, therefore, two coupled differential equations whose solutions describe the evolution of the field $E_1$ from $E^I_1$ to $E^F_1$ and   $E_2$  from $E^I_2$ to $E^F_2$.  These equations are coupled by an atomic scattering term, though a change in notation decouples them.   Dropping the coordinate arguments for notational ease, we write for the initial fields  
\bea
E^{I}_1 &=& E^{I}_{a1} + E^{I}_{b1},\\
E^{I}_2 &=& E^{I}_{a2} + E^{I}_{b2},
\eea where 
\bea
&&E^I_{a1} = E^I_{a2} = i \mathcal{E}^I \exp{(i k_\parallel y)} \sin(k_\perp z),\\
&&E^I_{b1} = -E^I_{b2} = \mathcal{E}^I \exp{(i k_\parallel y)} \cos(k_\perp z).
\eea  By contrast,  
\bea
&&E^F_{a1} = E^F_{a2} = i \mathcal{E}^F \exp{(i k_\parallel y)} \sin(k_\perp z),\\
&&E^F_{b1} = -E^F_{b2} = \mathcal{E}^I \exp{(i k_\parallel y)} \cos(k_\perp z). 
\eea
  Note that the $E^F_{b}$'s are  proportional to $\mathcal{E}^I$ since they are not attenuated by the atoms.   This is because the $E^I_{b}$ field components cancel in the interference region.  The intensity of the standing wave before the atoms is simply 
 \be
 I^{I} = |E^I_{a1}+E^I_{a2}|^2 = |2E^I_{a1}|^2 = 4|\mathcal{E}^I|^2\sin^2(k_\perp z),
\ee 
while  after the atoms it is 
\be
I^{F} = |E^F_{a1}+E^F_{a2}|^2 =  4|\mathcal{E}^F|^2\sin^2(k_\perp z).
\ee

We now relate the intensities immediately before and after the atomic position to the intensities $P^I_{1,2}$ (without atoms) and $P^F_{1,2}$ (with atoms) at the imaging system where the fields $\vec{E}_1$ and $\vec{E}_2$ no longer overlap.   We now  focus our attention on the mirror image, the case for the real image is analogous.   Because $E^I_{a1}$ and $E^I_{b1}$ are $\pi/2$ out of phase, 
\be
P^I_1 = |E^I_{a1}|^2+|E^I_{b1}|^2 = |\mathcal{E}^I|^2,
\ee while 
\be
P^F_1 = |E^F_{a1}|^2+|E^F_{b1}|^2 = |\mathcal{E}^F|^2\sin^2(k_\perp z) + |\mathcal{E}^I|^2\cos^2(k_\perp z).
\ee 
 By eliminating $\mathcal{E}_I$ and $\mathcal{E}_F$, we arrive at the relations
\bea
I^I_1 &=& 4P^I_1\sin^2(k_\perp z), \\
I^F_1 &=& 4P^F_1 - 4P^I_1\sin^2(k_\perp z),
\eea
and the solution to the modified Beer's law in terms of the measured intensities becomes
\bea\label{newBeers}
 n \sigma_0   &=&-\ln \left [\frac{P^F_1}{P^I_1}  + \left (\frac{\sin^2(k_\perp z)-1}{\sin^2(k_\perp z)} \right ) \frac{P^I_1 - P^F_1}{P^I_1}  \right ]  \nonumber \\
 &&+ 4 \frac{P^I_1-P^F_1}{I_{sat}} \approx4 \frac{P^I_1-P^F_1}{I_{sat}}.
\eea

The last relation is valid under the assumption that we are operating in the high-intensity regime defined by $\text{min}(I^F) \gg I_{sat}$.    As discussed above, this provides larger dynamic range in the photon detection, while also providing the ability to accurately determine atom density in the presence of a finite imaging resolution due to the linearity of this expression.    Whether this criterion applies depends on the distance of the BEC from the sample since atomic clouds near the nodes of the standing wave experience lower incident probe intensities.  That this criterion is satisfied is ensured by our short TOF which places the BEC close-enough to the first antinode of the standing wave below the mirrored sample for $\text{min}(I^F) \gg I_{sat}$, while also providing a nearly in situ measurement of the atomic density.   Satisfying the criterion also depends on the peak column density, because for  high-OD gases it is possible for $I^F< I_{sat}$   even if  $I^I \gg I_{sat}$.  While our quasi-1D BEC has a high OD at short TOF, care is taken to employ sufficiently high probe intensity to satisfy the high-intensity criterion for all ODs encountered.
 
 \begin{figure}[t!]
 \includegraphics[width=0.35\textwidth]{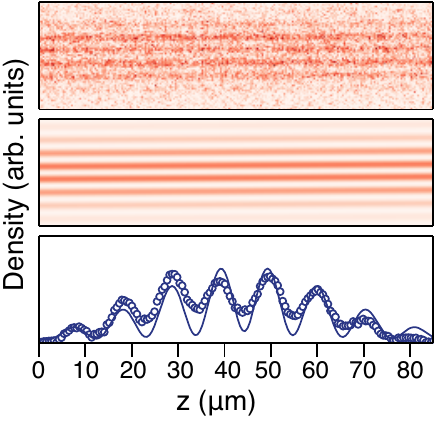}
\caption{Reflection angle measurement.   Top: absorption image of a thermal gas in the beam's interference pattern. Middle:  simulation of interference pattern. Bottom:  the imaging beam  angle $\theta$ is measured via a fit of the simulation's fringe periodicity to the data.}     
\label{Angle}
\end{figure}
 
\begin{figure}[t!]
\includegraphics[width=0.35\textwidth]{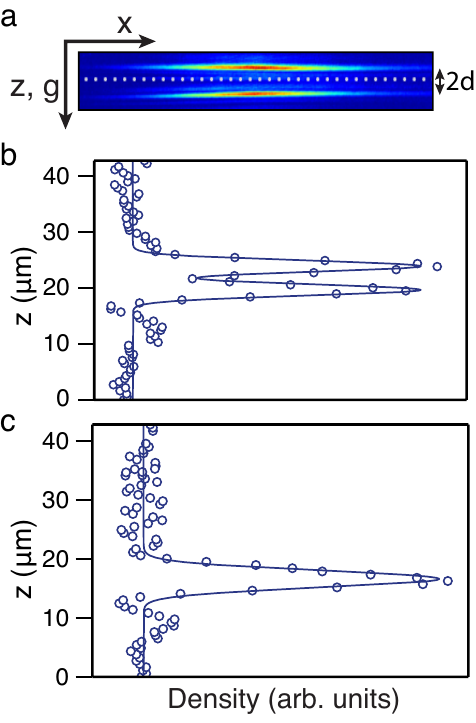}
\caption{ Calibration of distance below sample. (a) Example of in situ absorption image showing the real (lower) and mirror (upper) images equidistant from the mirrored sample surface (demarcated by white dots). (b) Example of a measurement of the distance of the atoms from the surface $d$.  The density has been integrated along $\hat{x}$ of an image such as that shown in panel (a). The width and positions of the double Gaussian profiles are left as free variables, resulting in  $d = 4.24(6)$~$\mu$m. (c) Fit to BEC close to the sample, using fixed BEC width, resulting in $d = 0.8(1)$~$\mu$m.}
\label{Distance}
\end{figure}

\subsection{Distance from sample}
All distance measurements rely on knowing the angle $\theta$ that the imaging beam makes with the sample. The periodicity of the interference pattern is measured by scattering the reflected beam off a large gas of thermal atoms located within the interference region. The resulting periodic modulation of the absorption of the atoms reveals the intensity modulation of the inference fringes.  See Fig.~\ref{Angle}.  A sinusoidal fit provides the periodicity of these  fringes $\lambda'$, which is related to the angle of incidence $\theta$ through $\theta= \arcsin{(\lambda/\lambda')}$, where $\lambda$ is the $^{87}$Rb D2-transition wavelength. The  angle for our measurements is $\theta=2.26(2)^\circ$. 

At this shallow angle of incidence, the distance $d$ of the gas from the sample surface is half the distance between the mirror and real images of the atoms.  See Figs.~\ref{Imaging} and~\ref{Distance}. Below $d\alt \sigma_\perp$   the two images merge, where $\sigma_\perp$ is the apparent transverse width of the gas (either physical or imaging resolution limited).
The height can be determined by fitting the images to two Gaussians of fixed width (known from images with $d> \sigma_\perp$) and choosing the center separation which best reproduces the observed total width. This results in a 10-fold increase in height uncertainty to 13\% when imaging gases below $d\approx1$~$\mu$m.  

Our data are taken at a short TOF such that the distance $d'$ to the sample at the time of imaging 
is closer to the first antinode of the imaging light standing wave and is in the $d'> \sigma_\perp$  regime of two discernible Gaussians. Fits to the resulting images provide an uncertainty on $d'$ of 60~nm. To find the uncertainty of $d$ we combine in quadrature this fit uncertainty with the uncertainty of inferring the height difference $\delta d$ between the in situ $d$ and $d'$.  We determine this error using two methods:  The first from repeated, sequential measurements of the atoms at $d$ and $d'$, and the second from the time needed for free-fall with an initial velocity imparted onto the atoms from the magnetic field gradients produced during trap release. The first method yields an error on $\delta d$ of 50~nm (60~nm) for the high-sensitivity (extended-dynamic-range) trap. For the second method, we need to measure the time of free fall $t$ and  the initial velocity $v$.  The time between trap release and imaging is $t=148(5)$~$\mu$s, which has contributions from the time needed to turn off the wire currents, the programmed TOF time, and the imaging duration.  The velocity is measured by a series images of the atoms just after release, and yields a $v = 5.5(3)$~(9.9(2))~$\mu$m/ms  for the high-sensitivity (extended-dynamic-range) trap.  Together these result in errors on $\delta d$ of 60~nm for both traps, roughly consistent with the first method. We assign an error of 60~nm to both traps' $\delta d$, resulting in a total error on $d$ of 80~nm.  

The microwires used to perform all measurements are 10~$\mu$m in length, with gaps in the gold-coated surface on either side. These gaps in the mirror surface result in shadows in the interference pattern above the sample.   However, this region extends only 180-nm below the region of the microwires, and therefore the shadow is not cast onto the gas trapped $\sim$1-$\mu$m below.  See Fig.~\ref{Imaging}.  The microwire arrays are spaced 300-$\mu$m apart so that the shadow from neighboring microwire arrays passes well below the BEC.

\begin{figure}[t]
 \includegraphics[width=0.48\textwidth]{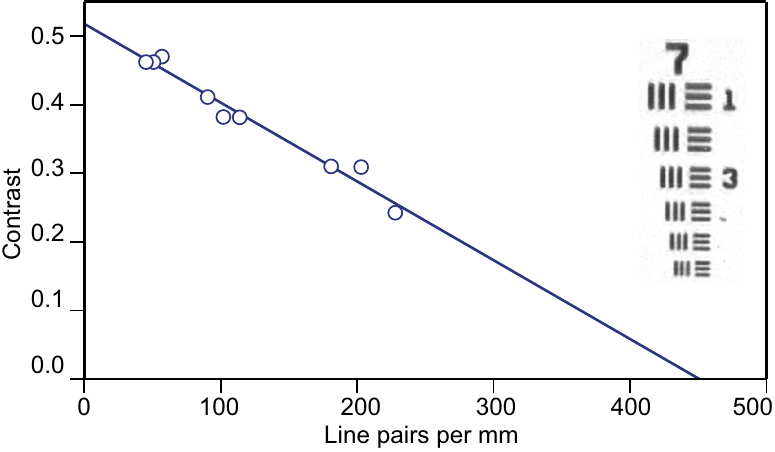}
\caption{Resolution measurement with test target.  Resolution measured through a viewport not under vacuum  using the 1951 Air Force Test Target, the image of which is shown in the inset.  The zero-contrast x-intercept indicates a resolution of  $2.2(1)$~$\mu$m.}
\label{Resolution_target}
\end{figure}

\begin{figure}[t]
 \includegraphics[width=0.48\textwidth]{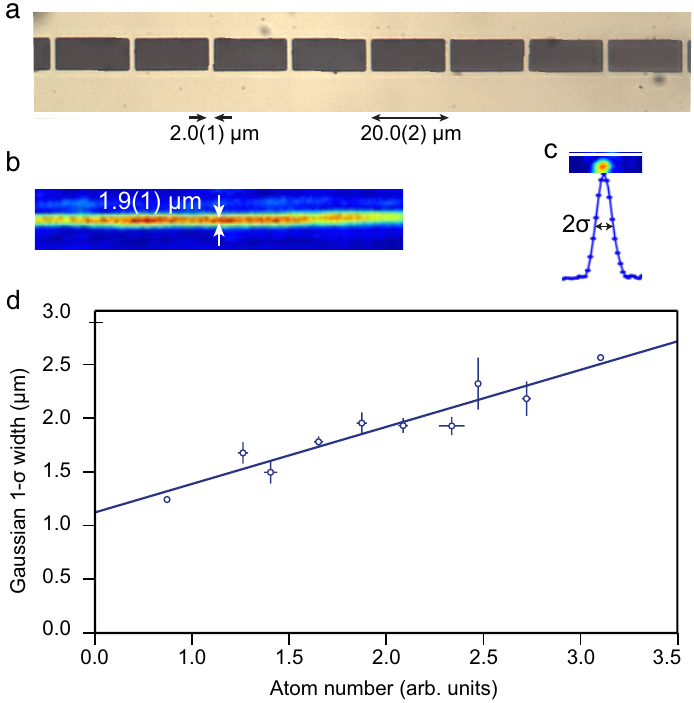}
\caption{Resolution measurement with 2-$\mu$m wire. (a) Image of 2.0(1)-$\mu$m microwires spaced 20.0(2)-$\mu$m apart. (b) Measurement of resolution in $\hat{z}$ using transverse width of unfragmented BEC (i.e., zero current flowing in calibration wire).  The FWHM, shown between the white arrows, is 1.9(1)~$\mu$m. (c) Image of one dot of the fragmented BEC in the potential of the microwires.  A Gaussian is fit to determine spot size; the $2\sigma$-width is indicated by the black arrows.  (d) Resolution versus atom number as measured by  fits such as that in panel (c). The y-intercept is $1.1(1)$~$\mu$m. }
\label{Resolution_2um}
\end{figure}

\begin{figure}[t]
 \includegraphics[width=0.48\textwidth]{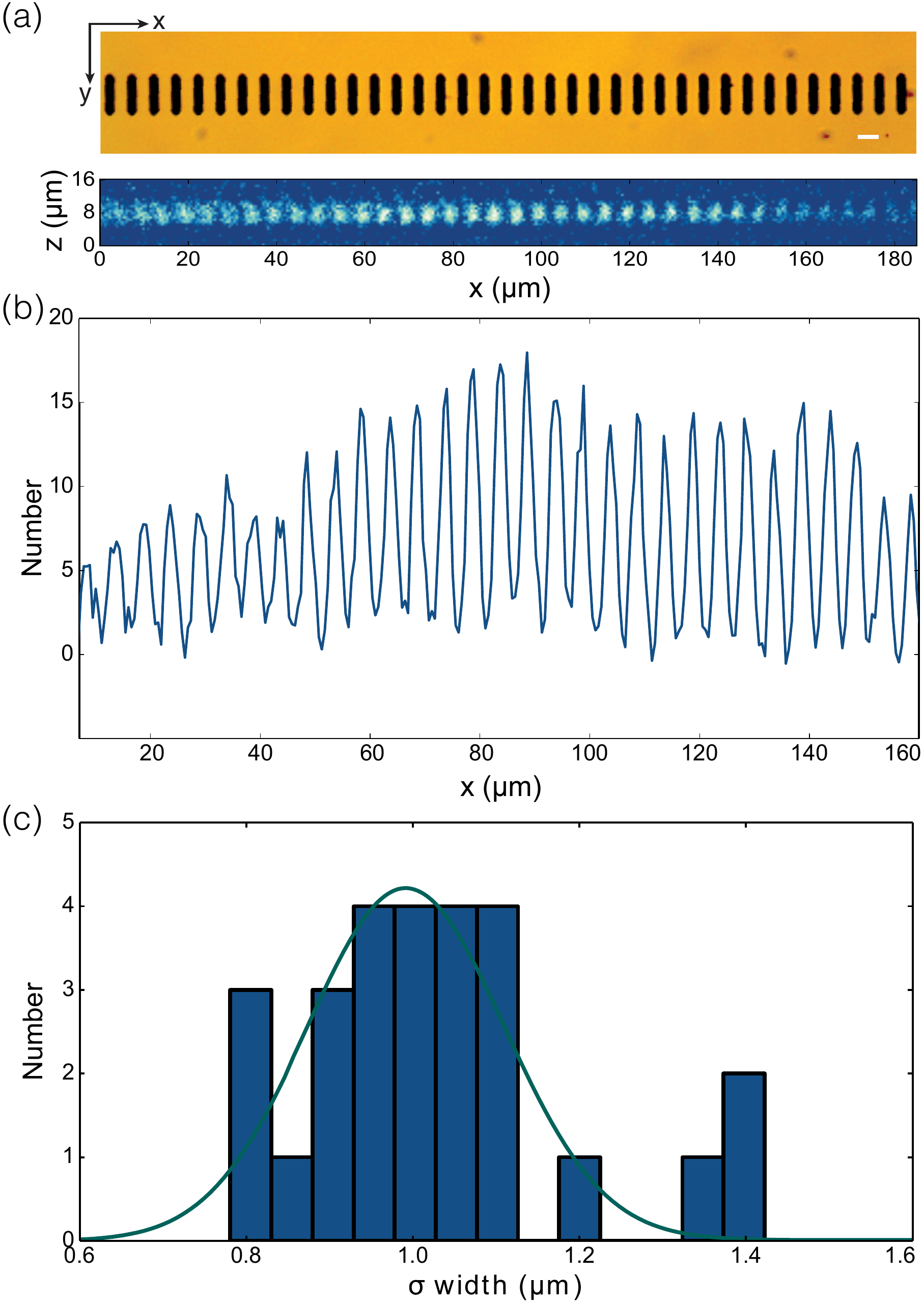}
\caption{Resolution measurements with array of 2.5-$\mu$m wires.  (a) Microwire array used for resolution measurement.  Wires are 2.5-$\mu$m wide, 2.5-$\mu$m spaced, and 10-$\mu$m long.   (b) Number of atoms per pixel along $\hat{x}$ integrated along the $\hat{z}$ direction. (c) Result of Gaussian fit to each peak in panel (b).  The histogram displays the 1-$\sigma$ widths of each of these peaks, and the green Gaussian is a fit to this distribution of widths.  The mean width is 1.0(1)~$\mu$m, providing a FWHM resolution estimate of 2.4(2)~$\mu$m.}
\label{Resolution_5um}
\end{figure}

\subsection{Resolution}
The first lens of the imaging system is an off-the-shelf 31.25-mm focal length, aspheric lens.  This is followed by a 24$\times$ telescope chosen such that each pixel of the CCD camera is 542(5)-nm wide in the object plane and well below our imaging resolution.  Future work will employ a custom lens system for improved resolution.

The resolution of the current imaging system is  measured in four ways.  First, we imaged the 1951 Air Force Test Target out of vacuum, but through an identical bucket window to the one used for the science chamber.  The window was tilted at $\sim$2.5$^\circ$ to approximate the angle of the absorption beam reflected off the mirrored sample.  Figure~\ref{Resolution_target} shows a measurement of contrast versus line pairs/mm, from which we determine a minimum observable line width of $2\times1.11(5)$~$\mu$m, and a line pair resolution of 2.2(1)~$\mu$m.    This is $\sim$20\% larger than the geometrically determined 1.8-$\mu$m diffraction limit (Rayleigh-criterion) of the lens system.  

We perform three \textit{in vacuo} measurements.  The first, shown in Fig.~\ref{Resolution_2um}(b), is a measurement of the transverse width in $\hat{z}$  of the quasi-1D BEC.  The 1-$\sigma$ width is 0.8(1)~$\mu$m, indicating a FWHM resolution of 1.9(1)~$\mu$m.  Assuming a cylindrically symmetric trap,  the resolution in $\hat{y}$ is at least $\sim$1.9(1)~$\mu$m for single-shot measurements, but from knowledge of trap parameters and density, we believe the $\hat{y}$-resolution is closer to the FWHM width of the BEC, 950~nm, convolved with $d$. (The stepping resolution the translation stage is 10~nm.)

The next two methods measure the in situ resolution in $\hat{x}$.  First we utilize a dimple trap formed by running current in the 2-$\mu$m-wide microwire array shown in Fig.~\ref{Resolution_2um}(a).  The BEC above these sparsely spaced wires fragments into a chain of dots, the width of one of which, shown in Fig.~\ref{Resolution_2um}(c), we measure as a function of atom number.  Figure~\ref{Resolution_2um}(d) shows the 1-$\sigma$ radius of the dot of Bose-condensed atoms as repulsive  mean field energy is reduced by trapping fewer atoms in the initial condensate.  The trap population is controlled by changing the RF evaporation cooling time.  Extrapolating to zero atoms, we obtain a Gaussian point-spread-function width of $2\sigma = 2.2(2)$~$\mu$m, for a FWHM resolution of 2.6(2)~$\mu$m.    

The last method, shown in Fig.~\ref{Resolution_5um}, uses an array of 2.5-$\mu$m-wide wires spaced 2.5-$\mu$m apart and finds the mean 1-$\sigma$ width of the dots of atoms in each of the microtraps.  The mean 1-$\sigma$ width is 1.0(1)~$\mu$m, implying a FWHM resolution of 2.4(2)~$\mu$m.  All of these methods are consistent with one another within less than 2$\sigma$.  We choose to take the adjusted weighted averages~\cite{Paule:1982} of the measurements to obtain the  SQCRAMscope atom density  and field FWHM resolvability of 2.2(1)~$\mu$m.  

The resolvability of current sources a distance $d$ away from the in situ gas position is reduced by the convolution of the field resolvability with the field propagation transfer function, i.e., the Biot-Savart Green's function to be described below.  For a wire with transverse dimensions much smaller than $d$, the current path resolvability is $\sqrt{\sigma_r^2+d^2}=2.3(1)$~$\mu$m for a BEC distance $d = 0.8(1)$~$\mu$m, or 2.8(1)~$\mu$m at $d = 1.7(1)$-$\mu$m away.

\section{Field sensing calibrations}\label{FieldCalib}

\subsection{Field-to-current mapping}
The mapping from measured field to inferred current distribution is performed using the procedure outlined in Ref.~\cite{Roth1989} and also employed in Refs.~\cite{Schmiedmayer05_Nature,Schmiedmayer06_APL,DellaPietra:2007ky,Aigner:2008}.  Using a Green's function that accounts for the finite thickness of the source wires, the Biot-Savart mapping between field and current is
\bea\label{Mapping}
j_y(k_x,k_y) &=& \frac{\bar{k}b_x(k_x,k_y,d)}{\mu_0 \sinh(\bar{k}h/2)} e^{\bar{k}(d+h/2)},\label{Mapping1}\\ 
j_x(k_x,k_y) &=&-\frac{k_y}{k_x}j_y(k_x,k_y), \label{Mapping2}
\eea
where $j_i$ is the current density in a wire of thickness $h$, and $\bar{k}=\sqrt{k_x^2+k_y^2}$ is the spatial wavenumber.  The measured $B_x(x,y)$-field map taken at $d$ is first Fourier-transformed with a spatial FFT algorithm.  Then,  $b_x(k_x,k_y,d)$ is filtered with the Green's function effecting the Biot-Savart mapping from field-to-source current at a distance $d$, i.e., applying Eq.~\ref{Mapping1}.  A Hanning window is numerically applied to the current density $j_y(k_x,k_y)$ to remove spurious high spatial frequencies.  Equation~\ref{Mapping2} yields $j_x(k_x,k_y)$, and finally an inverse FFT algorithm provides $J_x(x,y)$ and $J_y(x,y)$.
 \begin{figure}[t!]
 \includegraphics[width=0.48\textwidth]{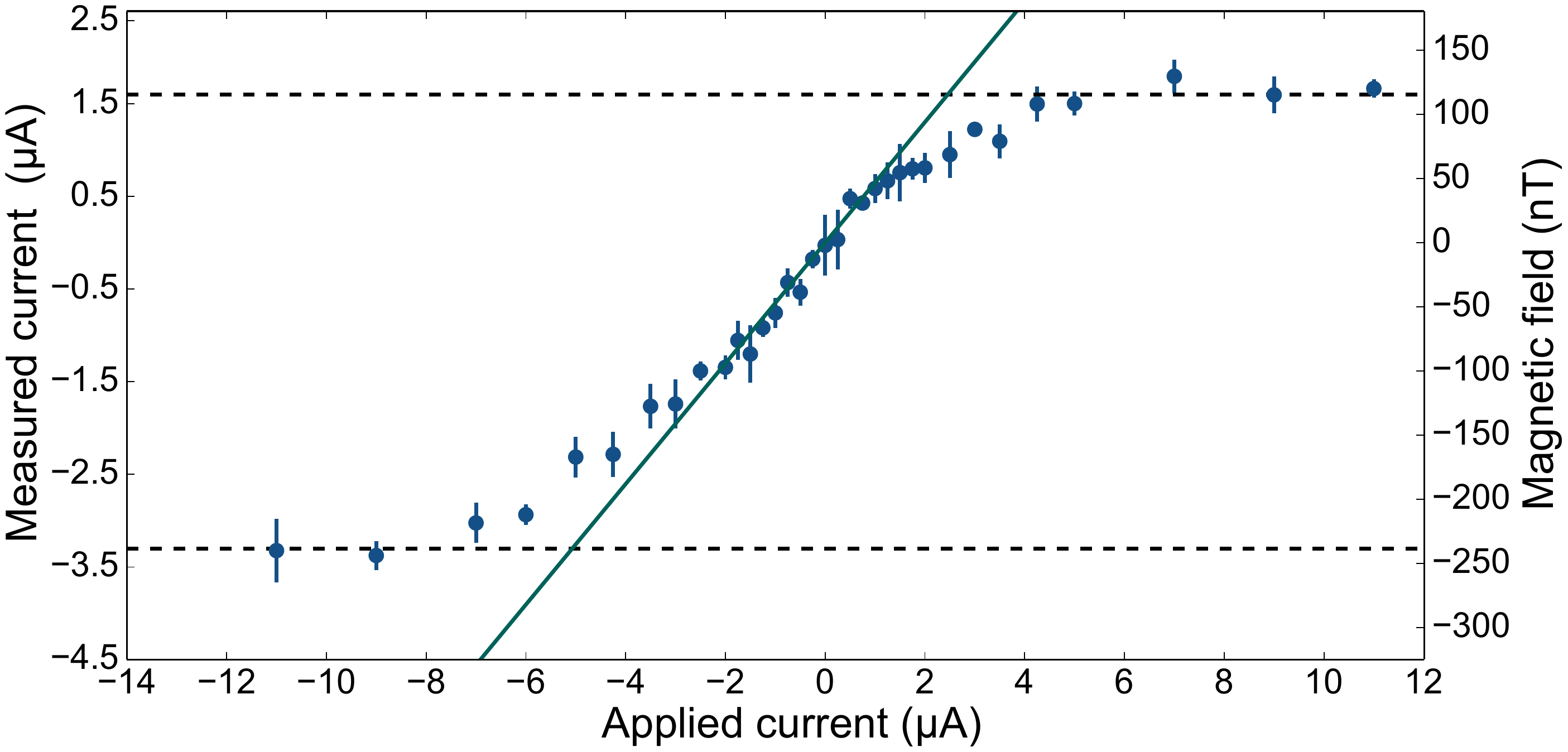}
\caption{Accuracy of extended-dynamic-range trap.   Measurement of current in the 6-$\mu$m-wide gold calibration  wire.   BEC is positioned 1.3(4)~$\mu$m below  wire center using the extended-dynamic-range trap, and the current is varied to create a density dimple or peak in the atomic density. The green line is a fit to the linear region.}
\label{EDRaccuracy}
\end{figure}

\begin{figure}[t]
 \includegraphics[width=0.48\textwidth]{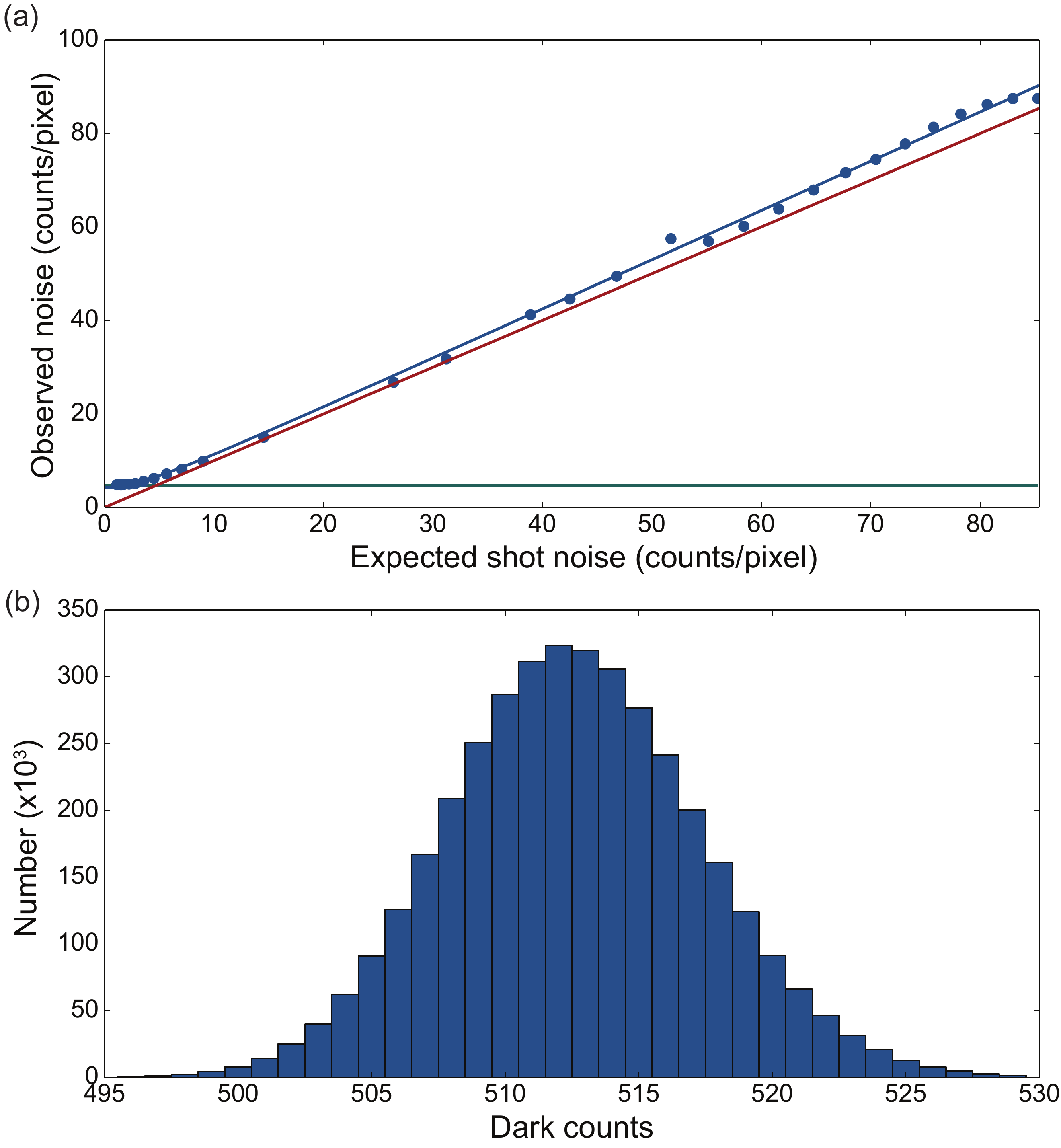}
\caption{Imaging noise measurement.  (a) Observed noise-per-pixel versus expected photon-shot-noise-per-pixel. Below $\sim$3600 counts-per-pixel (50 expected noise counts), the noise (blue data) is nearly at the expected photon-shot-noise-limit (red line).  But below $\sim$25 counts-per-pixel (5 expected noise counts), the noise is dominated by readout noise of the CCD. The green line  shows the contribution due to these fluctuations. We measure this level using the reference dark images taken during absorption imaging. The blue line is a fit to the data. (b) Histogram  of photon counts in each pixel of the dark image, with a Gaussian fit of $\sigma=4.68(2)$ counts from the readout noise. 
}
\label{PhotonNoise}
\end{figure}

\subsection{Accuracy of extended-dynamic-range trap}

We measured the accuracy, linearity, and dynamic range of the extended-dynamic-range trap in a manner similar to that presented in Fig.~\ref{fig3accuracy}.  The data are plotted in Fig.~\ref{EDRaccuracy}, where the green line is fit to the linear region and has slope 0.65(5). This implies a 3$\times$-worse accuracy than the high-sensitivity trap, possibly because it operates closer to the limit of the quasicondensate regime.  As for the data in Fig.~\ref{fig3accuracy}, we use a calibrated current source for generating the microwire current.  From this slope  we find that the responsivity is 35(5)\% lower than the expected $R=4.81(8)$~nT/(atom/$\mu$m) for this trap.

\subsection{Imaging noise}
Noise contributions in the three images arise from the noise of the resonant laser beam  as well as electronic readout and dark counts from the CCD camera.  We measure the per-pixel noise by performing absorption imaging without atoms present.  We care about the noise in a single image, and so rather than record the shot-by-shot statistics of counts in one pixel,  which would include an irrelevant contribution from laser intensity noise, we use the inhomogeneous imaging light intensity across the pixel array to measure the statistics of a single-shot image. Each pixel of the  image  is binned  by mean photon number, and  the variance versus mean of the number of counts in each bin is plotted in Fig.~\ref{PhotonNoise}a.  The line of unity slope represents the photon shot-noise-limit. For short exposure times, the chilled CCD camera exhibits only 0.07 counts/pixel/s of dark count noise, but readout contributes $\sigma=4.59(2)$ counts/pixel during our imaging time, as measured by examining the count statistics of the  dark image taken during absorption imaging; see  Fig.~\ref{PhotonNoise}(b). This readout noise dominates over shot noise below $\sim$5 counts per pixel. We observe that the imaging system is shot-noise-limited between approximately 100--3600 counts-per-pixel. The high-sensitivity trap is operated at 3000 counts-per-pixel, contributing to an equivalent field noise of 2.5(4)~nT, which is consistent with the expected photon shot noise of 2.2~nT.  In order to stay within the strongly saturated regime of atomic scattering, the extended-dynamic-range trap, which confines atoms at higher peak atomic density, is operated at 4500 counts-per-pixel.

\subsection{Atom noise}
\begin{figure}[t]
 \includegraphics[width=0.48\textwidth]{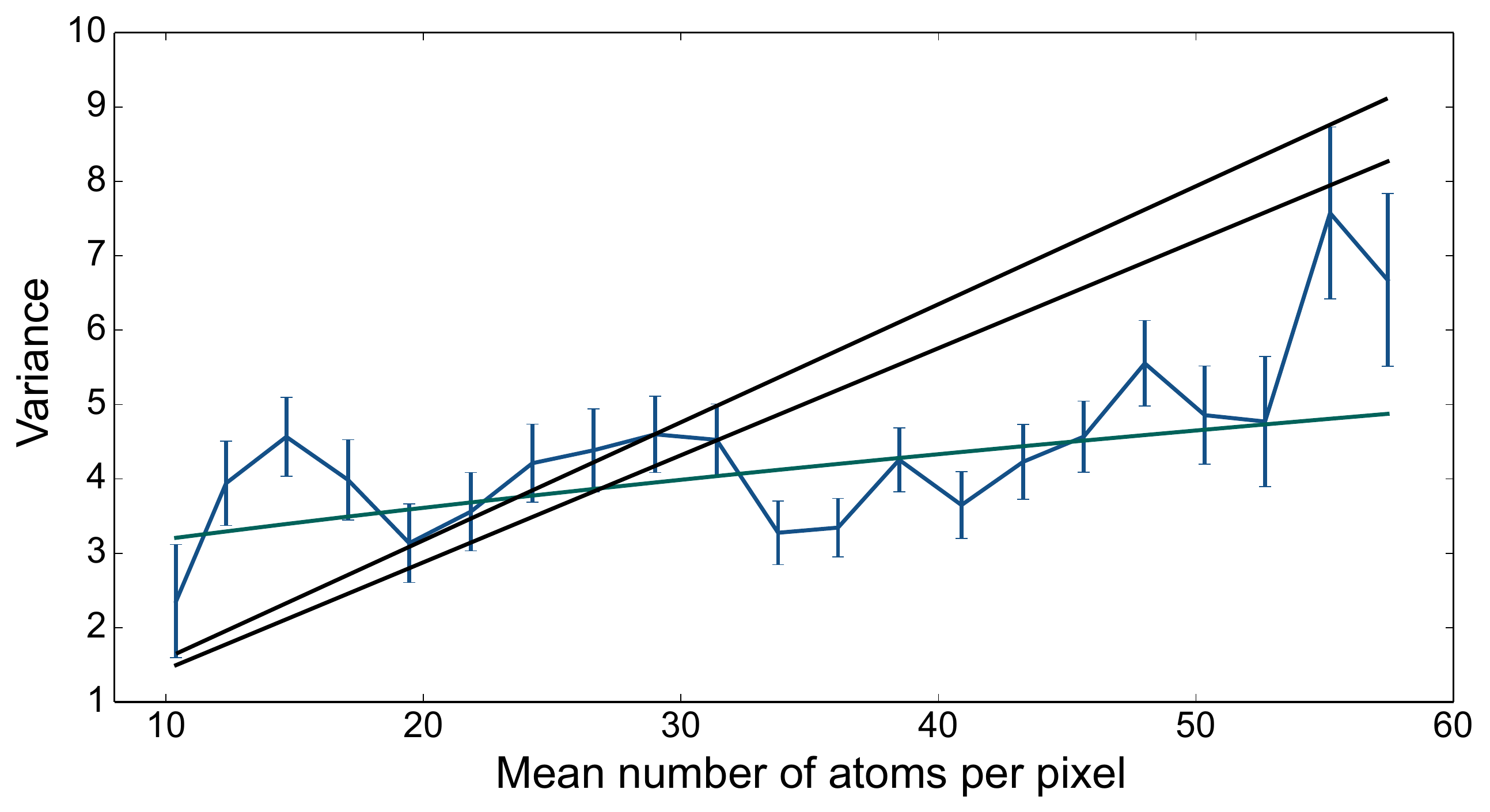}
\caption{Atom noise measurement. Variance versus mean-atom-number-per-pixel in images of  quasi-1D BEC density. The data, with photon shot noise subtracted, are fit to expressions from Ref.~\cite{gerbier} and  result in the  green curve.  The black lines correspond to the atom shot-noise-limit (unity slope) for imagining resolutions of 2.2~$\mu$m (top  black line) and 2.5~$\mu$m (bottom black line).} 
\label{AtomNoise}
\end{figure}

Images of the BECs contain atomic density noise in addition to photon shot noise, and the atomic density noise can come from both shot-to-shot variance in the total trapped atom number as well as intrinsic atomic density fluctuations from position-to-position along the quasicondensate.   The  variance in total atom number is recorded  in each shot and accurately accounted for in each run's density--to--field mapping as long as the quasicondensate equation of state remains valid.  However, the pixel-by-pixel density fluctuation does contribute to the noise floor.  To eliminate the contribution of total trap population fluctuations to the measurement of the intrinsic atom density noise, we compare the pixel-by-pixel density of a single-shot image to the mean density in each pixel expected for the gas, where we define a pixel to mean the $\hat{z}$-integrated atomic density.  This is accomplished by fitting the imaged density profile to a  quasicondensate profile~\cite{gerbier} for the same total atom number and trap parameters. This provides a mean and residual atom number for each pixel of this image, and we repeat this procedure for many such images.  We then bin the means and find the residual variance for each mean.  These are plotted in Fig.~\ref{AtomNoise}, where the black lines are the atom shot-noise limits for two imaging resolutions. The green line is the fit to the data using expressions from Ref.~\cite{gerbier} for a quasicondensate.  The fit allows us to measure the temperature of the gas~\cite{gerbier,jacqmin}, which is otherwise difficult to do given the short TOF permitted by our imaging system and the small number of atoms in the thermal wings at this temperature.

We see that the atom noise is sub-Poissonian for atom numbers-per-pixel above $30$.  The sub-shot-noise atom noise statistics are indicative of a quasicondensate, as discussed in Ref.~\cite{jacqmin}.  We operate the experiment such that the majority of pixels contain atoms in the 30--50 per-pixel range.  Mean atom numbers below 10 correspond to regions of the gas not described by the quasi-1D equation of state, either because these are thermal atoms pushed out from the center of the gas in to the wings~\cite{griffin2009bose} and/or because this is the region in which the validity of Thomas-Fermi approximation breaks down.  As such, these regions are not used for magnetometry, while numbers above $\sim$57 correspond to small regions around the centers of unusually dense BECs, leading to worse statistics in the figure. The average atom noise is 1.7(7)~nT, slightly less than that from photon shot noise.  

The atom shot-noise limits differ from  unity slope due to the artificial averaging of fluctuations~\cite{Hung:2011hy}.  This is caused by the blurring due to finite imaging resolution, expansion during the short TOF, and atom diffusion arising from the random recoil imparted during imaging.   The diffusion spot size is known to be accurately modeled by the diffusion equation $r_\text{RMS} = \sqrt{Nv^2_r\Delta t^2/3}\propto t^{3/2}$, where $N$ is the number of photons scattered and $v_r=5.88$ mm/s is the recoil velocity  on the imaging transition~\cite{ketterle:1999de}. The value of $r_\text{RMS}$ is $\sim$530~nm for our parameters. This is included in the measurements of the imaging resolution. Accounting for these effects, we obtain the two shot-noise limits in Fig.~\ref{AtomNoise} associated with imaging resolutions between 2.2--2.5~$\mu$m.

\begin{figure}[t!]
\includegraphics[width=0.48\textwidth]{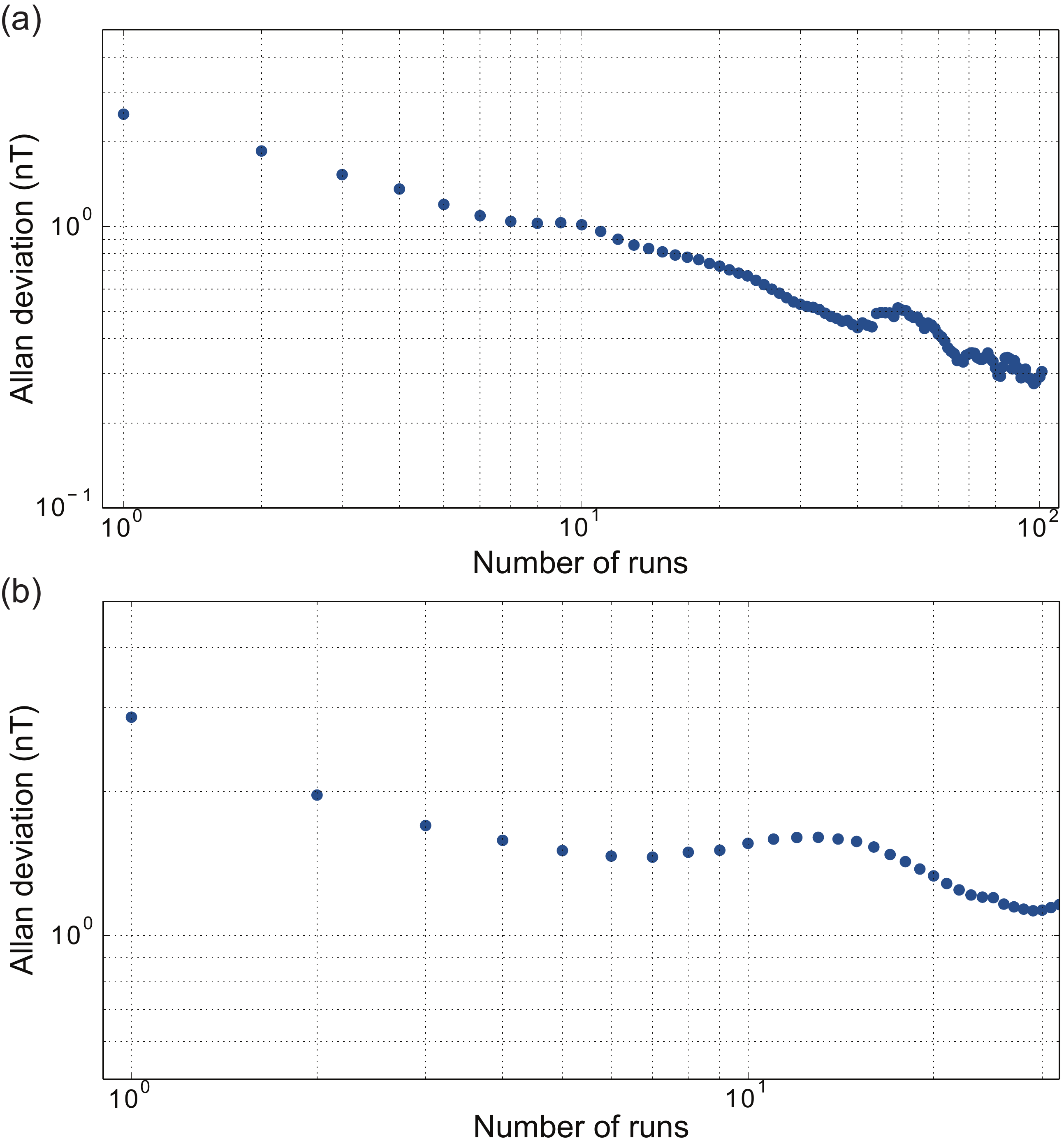}
\caption{Magnetometry stability.  (a) Allan deviation of the minimum detectable field. Noise floor is $\sim$300~pT per pixel (no spatial averaging) after 100 runs of the experiment.  (b) Allan deviation of the  repeatability of measurements of the 30-nT field generated by current running though the 12-$\mu$m-wide wire used in the accuracy measurements presented in Fig.~\ref{fig3accuracy}. The stability in the repeatability of the field is $\sim$1.1~nT after 30 runs.}
\label{AllanDev}
\end{figure}

\subsection{Allan deviation}

We measure the stability of the magnetometer using two methods.  In Fig.~\ref{AllanDev}a, we present a  measurement of the Allan deviation of the minimum detectable field above the same 12-$\mu$m-wide wire used for the accuracy measurements in Fig.~\ref{fig3accuracy} without current running in the wire. This field has a noise floor of $\sim$300~pT per pixel (no averaging) after 100 runs of the experiment, which takes roughly half an hour, though we collect information on $\sim$100 pixels during this time and thus spend less than 20~s per pixel. In the second measurement, we investigate the  Allan deviation of the  repeatability of the measurements of the 30-nT field  created by flowing 750~nA through the 12-$\mu$m wide wire with the atoms trapped 1.2-$\mu$m above.   The stability of the field is $\sim$1.1~nT after 30 runs. 
 Factors that may contribute to this minimum stability level are the residual sloshing of the motion of the BEC within the trap and the drift of the position of the BEC relative to the sample.  

\begin{figure}[t!]
 \includegraphics[width=0.48\textwidth]{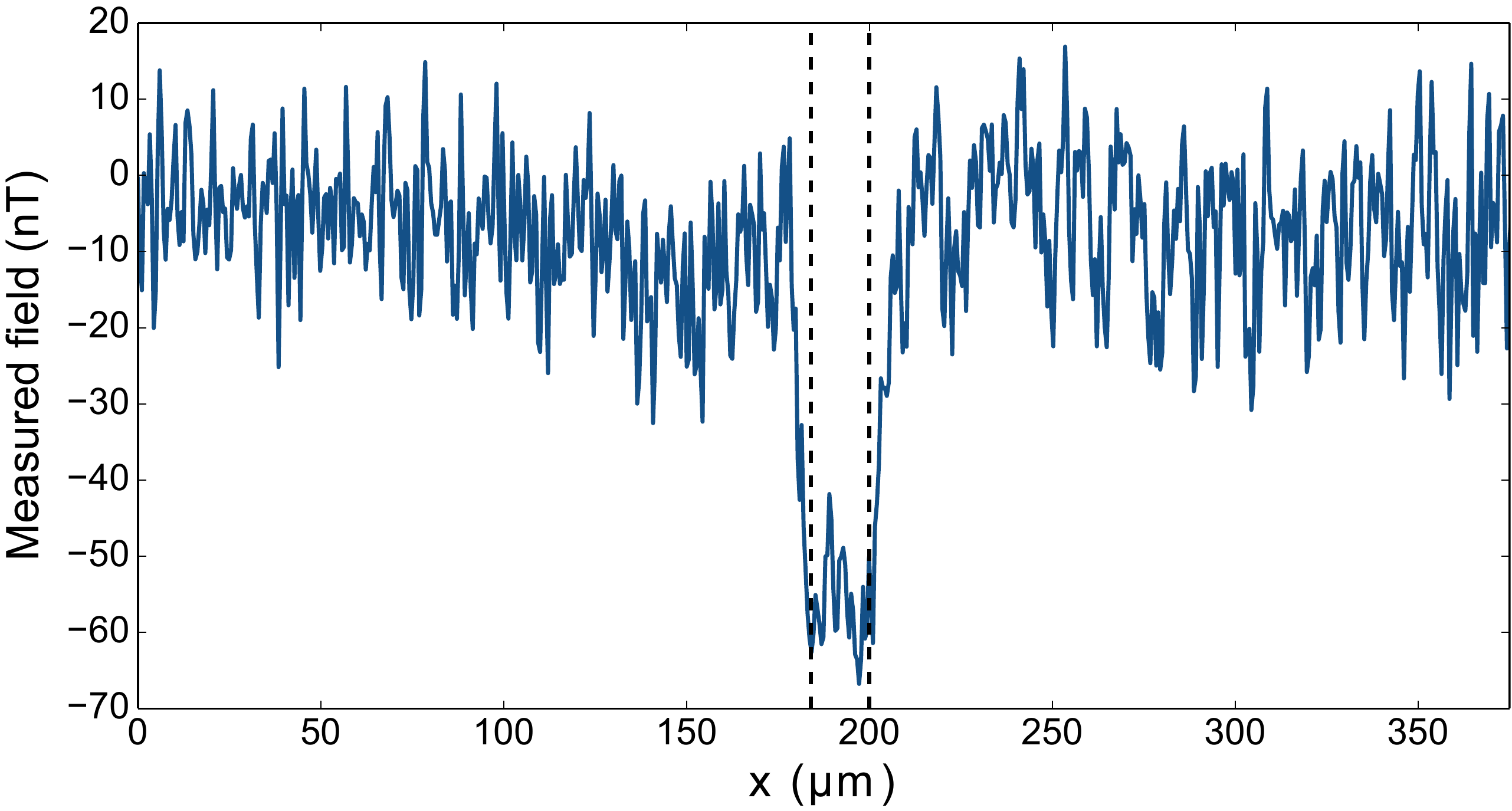}
\caption{Patch field measurement. 
Measured fictitious magnetic field 1.4(3)-$\mu$m from the sample due to a patch potential on the 16-$\mu$m microwire. This potential arises  after a BEC has been created $\sim$1,000 times within a few microns below the array. Black dashed lines indicate the positions of the wire edges.}
\label{Patch}
\end{figure}

\subsection{Patch fields}
Our SQCRAMscope is  sensitive to other forces from the sample surface besides magnetic fields.  With BECs trapped in atom chips near surfaces, Casimir-Polder forces have been detected~\cite{Lin:2004eya,Obrecht:2007ep} and used to determine the position of carbon nanotubes~\cite{Gierling:2011cl}.  Electric fields from surface charges can be detected~\cite{Schmiedmayer06_APL}. Such fields may arise from adsorbed Rb atoms, which can pose a difficulty for various applications~\cite{mcguirk,obrecht,Hattermann:2012ho,Carter:2012jb,Chan:2014kb,Fan:2015ia,Thiele:2015cr,Sedlacek:2016fl,Naber:2016jw}.  The electric field is due to an induced dipole whose magnitude varies depending on surface properties.  The force increases as more atoms are absorbed onto the surface, which may occur as atoms escape the atom chip trap during loading, RF evaporation, or during TOF expansion. 

Indeed, we observe an electric force on the atoms once we have created a BEC below a particular place below the calibration chip more than a thousand times.  The effects vanish once we move the BEC a few tens of microns to a fresh location.   Figure~\ref{Patch} shows this effect after $\sim$1000 repetitions with a BEC positioned $\sim$1~$\mu$m from a room-temperature, 16-$\mu$m-wide gold wire.  The wire is 400-nm taller than the surrounding silicon substrate, and so the electric field from the adatoms on the surface of the wire dominates that from the substrate to either side.  

The electric field from the adsorbed atoms creates a local potential above the wire. This is due to several effects: 1) The closer proximity of the Au to the atoms than the Si; 2) The different polarizabilities of Rb on Au versus Si; and 3) The potentially different adhesion characteristics of Rb to Au vs Si. This potential reduces the magnetometer's dynamic range and increases its sensitivity to sample height fluctuations.  While the per-atom Rb electric dipole has been measured for several materials~\cite{mcguirk,obrecht}, we are unaware of any measurement for Au. 

We do not expect this patch-field effect play a deleterious role in future SQCRAMscope magnetometry experiments because: 1) Most samples will have a smooth surface of a homogeneous material, reducing the force on the atoms from electric field gradients to a negligible level;   2) Samples can be on the mm-scale, allowing the use of fresh portions of the sample when needed; and 3) Fresh samples may be introduced with minimal downtime.  Nevertheless, we can investigate the efficacy of various methods for adatom removal that have been tried by other groups, including  sample heating, laser ablation and UV-light desorption~\cite{mcguirk,obrecht,Hattermann:2012ho,Carter:2012jb,Chan:2014kb,Fan:2015ia,Thiele:2015cr,Sedlacek:2016fl,Naber:2016jw}.


%

\end{document}